Astronomy
&
Astrophysics

# How the super-Eddington regime regulates black hole growth in high-redshift galaxies

Warren Massonneau[1], Marta Volonteri[1], Yohan Dubois[1], and Ricarda S. Beckmann[1,2]

[1] Institut d'Astrophysique de Paris, CNRS, Sorbonne Université, UMR7095, 98bis Bd Arago, 75014 Paris, France
e-mail: warren.massonneau@iap.fr
[2] Institute of Astronomy and Kavli Institute for Cosmology, University of Cambridge, Madingley Road, Cambridge CB3 0HA, UK

**ABSTRACT**

Super-Eddington accretion is one scenario that may explain the rapid assembly of $\sim 10^9\,M_\odot$ supermassive black holes (BHs) within the first billion year of the Universe. This critical regime is associated with radiatively inefficient accretion and accompanied by powerful outflows in the form of winds and jets. By means of hydrodynamical simulations of BH evolution in an isolated galaxy and its host halo with 12 pc resolution, we investigate how super-Eddington feedback affects the mass growth of the BH. It is shown that super-Eddington feedback efficiently prevents BH growth within a few Myr. The super-Eddington accretion events remain relatively mild with typical rates of about 2–3 times the Eddington limit, because of the efficient regulation by jets in that regime. We find that these jets are powerful enough to eject gas from the centre of the host galaxy all the way up to galactic scales at a few kpc, but do not significantly impact gas inflows at those large scales. By varying the jet feedback efficiency, we find that weaker super-Eddington jets allow for more significant BH growth through more frequent episodes of super-Eddington accretion. We conclude that effective super-Eddington growth is possible, as we find that simulations with weak jet feedback efficiencies provide a slightly larger BH mass evolution over long periods of time ($\sim 80$ Myr) than that for a BH accreting at the Eddington limit.

## 1. Introduction

Over 200 quasars at high redshift, $z \geq 6$ (i.e. less than $\sim 1$ Gyr after the Big Bang) have been discovered in the past decades (e.g. Fan et al. 2004; Mortlock et al. 2011; Wu et al. 2015; Bañados et al. 2018). These quasars host supermassive black holes (BHs) with masses larger than $M_{\rm BH} \geq 10^9\,M_\odot$. The formation and rapid growth of these compact objects at such an early stage of the Universe are one of the most important puzzles faced by modern astrophysics.

Several models have been suggested regarding the formation of seed BHs (e.g. Haiman 2013; Inayoshi et al. 2020; Volonteri et al. 2021, and references therein). There are several potential formation channels: light seeds, which are Pop III remnant BHs of $M_{\rm BH} \simeq 10^{1-2}\,M_\odot$ (e.g. McKee & Tan 2008; Hosokawa et al. 2011; Hirano et al. 2014, 2015; Stacy et al. 2016) formed in dark matter (DM) minihalos; intermediate seeds, corresponding to intermediate mass BHs born in dense clusters from stellar or BH mergers, with $M_{\rm BH} \simeq 10^3\,M_\odot$ (e.g. Devecchi & Volonteri 2009; Katz et al. 2015; Sakurai et al. 2017, 2019; Tagawa et al. 2020); heavy seeds, which emerge from the direct collapse of first halos and that leaves a seed with a large initial mass of $M_{\rm BH} \simeq 10^{4-5}\,M_\odot$ (e.g. Bromm & Loeb 2003; Inayoshi et al. 2014; Regan et al. 2014; Latif et al. 2016; Wise et al. 2019). Even for massive seeds, assuming an Eddington-limited accretion, the orders of magnitude between formation mass and the observed mass of the BHs in $z \geq 6$ quasars requires a very high duty cycle of accretion close to the Eddington limit.

Massive halos can provide gas to funnel to the most central regions of high-redshift galaxies and feed BHs through direct accretion of cold cosmic streams, later on replaced by clumpy gas accretion (Bournaud et al. 2011; Di Matteo et al. 2012, 2017; Dubois et al. 2012b). Even in the presence of large gas reservoirs, continuous accretion at Eddington levels is challenging to sustain, as feedback from gas accretion onto the BH seed can severely affect gas inflows and prevent efficient BH growth (e.g. Johnson & Bromm 2007; Alvarez et al. 2009; Milosavljević et al. 2009; Dubois et al. 2013). In addition, feedback from massive stars in the shallow potential well of low-mass galaxies is able to significantly suppress the availability of cold interstellar gas in the central regions of galaxies, therefore, quenching BH growth in low-mass galaxies (Dubois et al. 2015; Anglés-Alcázar et al. 2017; Bower et al. 2017; Habouzit et al. 2017; Prieto et al. 2017; Trebitsch et al. 2018; Hopkins et al. 2022). This effect, also associated to the wandering of BHs in shallow potential wells of low-mass galaxies (Bellovary et al. 2019; Pfister et al. 2019; Ma et al. 2021) might prevent BHs to grow until their host galaxy build up a massive and compact enough bulge component (Lapiner et al. 2021). Accretion rates larger than the Eddington limit could solve this timescale issue, but only if super-Eddington episodes can be sustained long or frequently enough to allow for significant mass growth (Volonteri et al. 2015; Pezzulli et al. 2017).

Breaking the Eddington limit is a natural manifestation of non-spherically symmetric accretion (Paczynski & Abramowicz 1982) and over the past decades, observational evidence supports super-Eddington accretion. Detections of X-ray binaries, such as SS 433 (e.g. Okuda 2002), or of ultra-Luminous X-ray sources that may harbor stellar-mass BHs (e.g. Winter et al. 2006) are considered to be signatures of accretion rates above the Eddington limit. Several observations point to the supermassive

BHs in narrow-line Seyfert-1 galaxies (e.g. Mineshige et al. 2000; Collin & Kawaguchi 2004; Du et al. 2014, 2018; Jin et al. 2017), as well as transient sources called tidal disruption events (e.g. Burrows et al. 2011; Lin et al. 2017, 2022), which are also candidates for super-Eddington accretors. Several theoretical models have been developed to explain this state of accretion, such as the slim disc model (Abramowicz et al. 1988; Sądowski 2009; Abramowicz & Fragile 2013). One of the most important features of this regime is the photon trapping effect (Katz 1977; Begelman 1978), in which radiated photons are advected to the BH instead of being diffused out of the disc. As a result, the radiative efficiency drops below its typical value of $\epsilon_r \sim 0.1$ expected in thin discs. However, there is yet no agreement on how efficient this process is, and on the exact value of the resulting radiative efficiency (e.g. Ohsuga et al. 2005; Jiang et al. 2014, 2019; McKinney et al. 2014, 2015; Sądowski et al. 2015). Semi-analytical models using the slim disc regime have shown that a light seed could grow from 100 $M_\odot$ to a few $10^9$ $M_\odot$ within 1 Gyr (Madau et al. 2014), thanks to reduced radiative efficiency. Other numerical works in modelling the accretion disc near the BH found that if gas inflows are funneled in the equatorial plane with the radiation and outflows escaping out from the poles, then super-Eddington accretion can be sustained (e.g. Inayoshi et al. 2016; Sugimura et al. 2017; Takeo et al. 2018, 2020; Toyouchi et al. 2019, 2021; Kitaki et al. 2021).

Although numerical simulations of accretion discs are extremely detailed and cover a wide range of physical processes, such small-scale numerical experiments do not explore regions beyond the pc-scale and cannot capture the dynamics of the gas feeding the BH and the impact from super-Eddington feedback at larger galactic scales. Using sub-pc resolution hydrodynamical simulations of circum-nuclear discs with BHs releasing energy with thermal input, Lupi et al. (2016) showed that stellar-mass BHs are able to grow to $\sim 10^4$ $M_\odot$ within 3 Myr. This is thanks to a combination of the very high density gas clumps formed due to disc fragmentation, and the radiative inefficient accretion from this super-critical regime. Their results indicate that several events of short periods of super-Eddington accretion on light seeds can sustain enough mass growth to explain the most massive BHs powering quasars at high redshift. However, super-Eddington accretion may be accompanied (if not dominated) by powerful jets (e.g. Sądowski & Narayan 2015; Narayan et al. 2017), which, added to the radiation, can have an impact on BH growth. Recently, Regan et al. (2019) investigated how super-Eddington bipolar jet outflows affect gas inflows in an atomic cooling halo in a cosmological setting. They were able to reach resolutions down to the BH gravitational sphere of influence and found that jets almost entirely shut down the inflow of gas towards the BH on small scales ($\simeq 0.1$ pc) and were not able to break out of the halo. The effective accretion rate was found to be $\sim 4$ orders of magnitude below the Eddington limit, thus suggesting that BHs are not able to grow effectively above the limit with strong jetted outflows. On the other hand, Takeo et al. (2020) performed radiation hydrodynamics (RHD) simulations combining radiative and mechanical feedback and found that gas in the polar region was completely evacuated but gas inflows from the equatorial region maintained super-Eddington accretion rates. The jets were able to pierce through the gas distribution without disrupting it. The discrepancy between these results may be linked to the gas geometry, however it is unclear if the axisymmetric disc configuration in Takeo et al. (2020) can be achieved assuming the turbulent nature of high-redshift galaxies.

In this paper, we explore whether a BH can sustain super-Eddington accretion when both wind and mechanical feedback work together. We study the impact that this critical regime has on BH growth and on the gas properties of the host galaxy, from star-forming clouds to galactic scales. To this end, we implement in the adaptive mesh refinement code RAMSES (Teyssier 2002) accretion and feedback processes above the Eddington limit, and perform several simulations of an individual galaxy in an isolated DM halo. We add a BH to the center of the emergent galaxy and conduct a survey on the parameter dependence regarding BH growth and overall impact on the gas inflows and outflows, varying the feedback efficiencies and modes of injection in the super-Eddington regime.

The outline of the paper is as follows. We describe our numerical methods in Sect. 2, including a summary of the already implemented active galactic nuclei (AGN) feedback models and the new application for super-Eddington accretion and feedback in RAMSES. In Sect. 3, we describe the set-up for our isolated DM halo simulations, in particular the initial conditions that lead to the formation of the galaxy and the addition of the BH. We then present our simulation results in Sect. 4 where we analyse the importance of the super-Eddington AGN feedback on star formation, BH growth and gas properties, varying its power and modes of injection. We discuss these results and finally conclude in Sect. 6.

## 2. Implementation of the super-Eddington regime

This paper showcases a number of hydrodynamical simulations of an isolated DM halo in which BHs have been evolved in the super-Eddington regime, all of them performed using the adaptive mesh refinement code RAMSES, which was adapted to allow for accretion and feedback processes above the Eddington limit, defined as:

$$L_{\rm Edd} \equiv \frac{4\pi G M_{\rm BH} m_{\rm p} c}{\sigma_{\rm T}}, \quad (1)$$

where $G$ is the gravitational constant, $M_{\rm BH}$ the BH mass, $m_{\rm p}$ the proton mass, $c$ the speed of light and $\sigma_{\rm T}$ the Thomson cross-section. In this Section we detail how the super-Eddington regime is implemented in RAMSES.

### 2.1. Accretion onto the BH

A BH is represented by a "sink" particle that can transfer mass, momentum and energy from and to the gas. The algorithm for gas accretion onto sink particles, first introduced by Krumholz et al. (2004) for grid codes, is described in detail in Dubois et al. (2012a) for its RAMSES implementation. We recall here the main ideas: the BH is manually placed with a given initial mass, velocity and spin, in the simulation at a certain point in time. Once set, it is able to move, following the local gravitational acceleration. At all times, we enforce the highest resolution level within a sphere of radius $4\Delta x$, where $\Delta x$ is the smallest cell size, to better resolve the forces near the BH. Since we cannot resolve accretion discs around BHs, as their size is below our resolution limit, we calculate the accretion rate on the BH using the Bondi-Hoyle-Lyttleton (BHL) formula (Hoyle & Lyttleton 1939; Bondi 1952):

$$\dot{M}_{\rm BHL} = \frac{4\pi G^2 M_{\rm BH}^2 \bar{\rho}}{(\bar{c}_{\rm s}^2 + \bar{v}_{\rm rel}^2)^{3/2}}, \quad (2)$$

where the averaged density $\bar{\rho}$, sound speed $\bar{c}_{\rm s}$ and relative velocity between the BH and the gas $\bar{v}_{\rm rel}$ are computed over a sphere of radius of $4\Delta x$ with contributions from each cell weighted by $w \propto \exp(-r^2/r_K^2)$, (Krumholz et al. 2004). The kernel radius $r_K$ depends on whether the BHL radius $r_{BHL} = GM_{BH}/(c_s^2 + v_{rel}^2)$ is resolved or not, where $c_s$ and $v_{rel}$ are the sound speed and relative velocity in the cell where the sink lies. It is defined as follows:

$$r_K = \begin{cases} \Delta x/4 & r_{BHL} < \Delta x/4, \\ r_{BHL} & \Delta x/4 \leq r_{BHL} \leq 2\Delta x, \\ 2\Delta x & r_{BHL} > 2\Delta x. \end{cases} \quad (3)$$

In order to prevent having cells completely depleted of gas at high accretion rates, a safety check allows for accretion of only up to 25% of the total mass available in each cell at each timestep of size $\Delta t$, i.e. $\dot{M}_{floor} = 0.25\rho\Delta x^3/\Delta t$, where $\rho$ is the density of the cell. All the accretion rates mentioned are subject to the same weight $w$, applied when gas is removed from the grid and accreted onto the sink particle. The accretion rate is calculated as $\dot{M}_{acc} = \min(\dot{M}_{BHL}, \dot{M}_{floor})$, which is always less than the mass contained in the kernel divided by the timestep. The mass contained in the kernel is used to both feed the BH and power kinetic feedback, with mass conservation enforced at all times.

To calculate the feedback energy released by the BH in the super-Eddington regime, the slim disc solution is used (Abramowicz et al. 1988; Sądowski 2009; Abramowicz & Fragile 2013). Despite having accretion rates that go well above the Eddington value, the disc emits a luminosity that is only slightly above the Eddington luminosity $L_{Edd}$ (see Eq. (1)). It has been shown in several theoretical works (Begelman 1978; Mineshige et al. 2000; Ohsuga et al. 2002; Madau et al. 2014; Inayoshi et al. 2020) that when $\dot{M}_{acc} > \dot{M}_{Edd}$, the luminosity $L \propto \log(\dot{M}_{acc})$ which means that the radiative efficiency $\epsilon_r$ decreases. This is due to the photon-trapping effect and it impacts the growth rate of the BH, since $\dot{M}_{BH} = (1 - \epsilon_r)\dot{M}_{acc}$. In order to estimate $\epsilon_r$, we use the fit from Madau et al. (2014, who themselves use the fit from Sądowski 2009) for the ratio of the bolometric luminosity-to-Eddington $f_{Edd} \equiv L/L_{Edd}$:

$$f_{Edd} = A(a)\left[\frac{0.985}{1.6\dot{M}_{Edd}/\dot{M}_{acc} + B(a)} + \frac{0.015}{1.6\dot{M}_{Edd}/\dot{M}_{acc} + C(a)}\right], \quad (4)$$

where $\dot{M}_{Edd}$ is the Eddington mass accretion rate defined as $\dot{M}_{Edd} \equiv 10L_{Edd}/c^2$, given an $\epsilon_r$ of 0.1; $a$ is the BH spin, and with spin-dependent functions $A$, $B$ and $C$ as:

$$A(a) = (0.9663 - 0.9292a)^{-0.5639}, \quad (5)$$

$$B(a) = (4.627 - 4.445a)^{-0.5524}, \quad (6)$$

$$C(a) = (827.3 - 718.1a)^{-0.7060}. \quad (7)$$

The effect of photon-trapping on the radiative efficiency is illustrated in Fig. 1. The different markers represent different spins, from no rotation (blue triangle) to maximum rotation, set at 0.998[1] (green diamond). We also show the BH spin $a = 0.7$ (orange cross) giving $\epsilon_r \simeq 0.1$ for small values of $f_{Edd}$. $\epsilon_r = 0.1$ is usually adopted in cosmological simulations as a constant radiative efficiency (e.g. Di Matteo et al. 2008; Booth & Schaye 2009; Dubois et al. 2012a, although see Dubois et al. 2014, 2021; Bustamante & Springel 2019 for spin-dependent radiative efficiencies). At a given accretion rate, the higher the spin, the higher the radiative efficiency. While our simulations can self-consistently evolve BH spins and, thus, associated radiative efficiencies, we prefer to simplify the problem by adopting a fixed spin value of $a = 0.7$ ($\epsilon_r = 0.1$ for $f_{Edd} < 1$) throughout the rest of this paper. In this study, accretion is not Eddington-limited unless specified, in which case the accretion rate is limited by $\dot{M}_{Edd}$.

---
[1] The normalized spin $a = 0.998 < 1$, is due to the photons emitted by the accretion disc and captured by the BH (Thorne 1974).

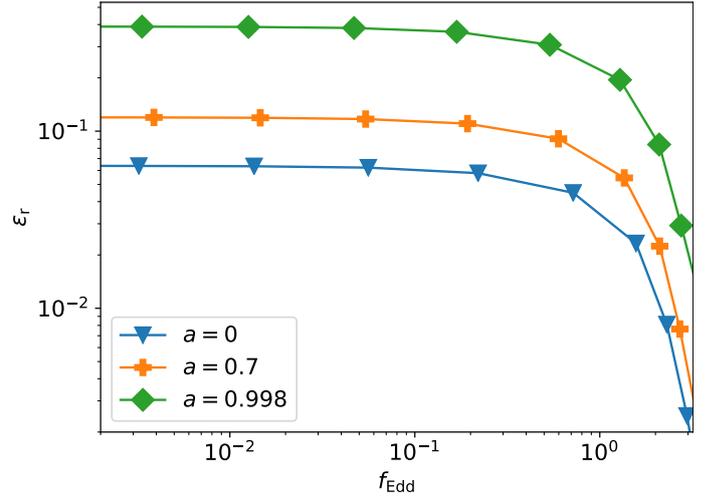

**Fig. 1.** Radiative efficiency $\epsilon_r$ as a function of the Eddington fraction $f_{Edd}$. The markers showcase the different BH spin values. Photon-trapping is visible when $f_{Edd} > 1$ for all spin parameters.

### 2.2. Feedback in the different regimes

In RAMSES, AGN feedback is implemented with two separate feedback modes: a kinetic mode, active at low accretion rates, which is modeled as a jet-like outflow and a thermal mode, active at high but sub-Eddington accretion rates, which heats the gas surrounding the BH by releasing thermal energy (Dubois et al. 2012a). For consistency with Madau et al. (2014) and Sądowski et al. (2016), the transition between these regimes is based on $f_{Edd}$ rather than on the ratio of accretion rates.

With our current understanding of the super-Eddington regime, super-Eddington AGN feedback includes both kinetic and radiative/thermal components. We have modified RAMSES to allow for a new super-Eddington feedback mode that injects both thermal and kinetic energy simultaneously.

#### 2.2.1. Kinetic feedback

In the sub-Eddington regime, when $f_{Edd} \leq 0.01$, feedback is kinetic, and modelled with jet-like outflows, called "radio" mode. At each feedback event, mass, momentum and energy are deposited in a cylinder of diameter and height equal to $2r_{jet}$. The size of the jet $r_{jet}$ is a user-defined parameter in units of $\Delta x$. The cylinder axis is aligned with the BH spin axis (Dubois et al. 2014). To calculate the energy injected as kinetic energy in all the cells within the cylinder, we follow Sądowski et al. (2016), which gives a total jet feedback of

$$\dot{E}_{jet} = \eta_{jet}\dot{M}_{BH}c^2, \quad (8)$$

where $\eta_{jet} = 1.3a^2 f_{MAD}^2$ is the efficiency factor of the kinetic feedback for a magnetically arrested disc (MAD) taken from Tchekhovskoy (2015), and $0 \leq f_{MAD} \leq 1$ represents the fraction of MAD strength (magnetic field saturation).

Since jets are a relativistic phenomenon and are launched at velocities close to $c$ on scales that cannot be resolved in our simulations (a few Schwarzschild radii), we must develop a subgrid model to account for how the jet evolves from the launching scale to the simulation resolution. As jets traverse the interstellar medium from the Schwarzschild scale to pc scales, they entrain mass and slow down (Chatterjee et al. 2019): the mass that loads the jet changes its velocity such that the jet velocity at the resolution scale is lower than that at the unresolved launching scale (that of the accretion disc). Since we cannot resolve the physical injection scale of the Schwarzschild radius, we inject the jet with a mass and velocity appropriate for the smallest resolved scale in the simulation, assuming kinetic energy has been conserved on intermediate scales. The mass-loading factor $\beta_{\rm jet}$ is the ratio between the mass ejection rate $\dot{M}_{\rm jet}$ in the jet at the resolution scale and the growth rate of the BH, i.e. $\beta_{\rm jet} = \dot{M}_{\rm jet}/\dot{M}_{\rm BH}$. Assuming that the jet energy at injection (Eq. (8)) is equal to the kinetic energy of the jet at larger scales, $\dot{E}_{\rm jet} = 0.5 \dot{M}_{\rm jet} v_{\rm jet}^2$, one can write the mass-loading factor as

$$\beta_{\rm jet} = 2\eta_{\rm jet}\left(\frac{v_{\rm jet}}{c}\right)^{-2}. \tag{9}$$

For our simulations we set $v_{\rm jet} = 0.1c$ to keep the time step of the simulations affordable, and therefore have $\beta_{\rm jet} = 200\eta_{\rm jet}$.

The extra mass that is loaded into the jet is transferred from the same cells that were used for the accretion, i.e. within $4\Delta x$ of the sink particle with the same distance-weighting described previously, from where it is redistributed to all cells enclosed within the jet (Dubois et al. 2012a). In the case where insufficient mass is available to fully load the jet, the mass-loading factor is smaller than the predicted value but the jet is still launched at $v_{\rm jet} = 0.1c$. A maximum of 25% of the total mass available in each cell is used to load the jet, to avoid dealing with extremely low densities and numerical instabilities.

### 2.2.2. Thermal feedback

When $0.01 < f_{\rm Edd} \le 1$, the AGN enters the so-called "quasar" mode, which corresponds to feedback coming from disc winds and radiation. In this mode, energy is released as thermal energy in a sphere of radius $r_{\rm thm}$ around the BH. This radius, similarly to $r_{\rm jet}$, is a user-defined parameter in units of $\Delta x$.

The efficiency factor $\eta_{\rm thm}$ for this regime corresponds to the thermal wind efficiency of the disc. Following Sądowski et al. (2016), the spin-dependent thermal wind efficiency can be written as $\eta_{\rm thm} = \eta_{\rm d \to thm} \times \eta_{\rm ISCO}$, with

$$\eta_{\rm ISCO} = 1 - \sqrt{1 - \frac{2}{3R_{\rm ISCO}(a)}}, \tag{10}$$

where $\eta_{\rm d \to thm}$ is the fraction of energy released from the disc coupled to the gas within $r_{\rm thm}$, depending on how well the energy from the accretion disc can be absorbed by the surrounding gas. We set $\eta_{\rm d \to thm} = 0.15$, which has been calibrated to reproduce the BH-galaxy scaling relations (Dubois et al. 2012a). $R_{\rm ISCO}$ corresponds to the innermost stable circular orbit (Bardeen 1970), i.e.

$$R_{\rm ISCO} = \frac{GM_{\rm BH}}{c^2}\left(3 + Z_2 \mp \sqrt{(3 - Z_1)(3 + Z_1 + 2Z_2)}\right), \tag{11}$$

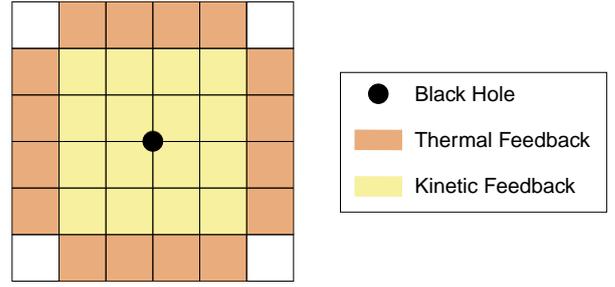

**Fig. 2.** Schematic representation of the feedback injection in cells around the BH in RAMSES. The region of injection of the kinetic feedback is in yellow and the thermal in vermilion. No thermal feedback is injected in the same region as the kinetic one, as the deposited energy is different for both forms.

where the $\mp$ sign corresponds to respectively the positive (−) and negative (+) values of the BH spin, and

$$Z_1 = 1 + (1-a^2)^{1/3} \times \left((1+a)^{1/3} + (1-a)^{1/3}\right),$$

$$Z_2 = \sqrt{(3a^2 + Z_1^2)}.$$

Including this effect, the total energy deposited in the cells contained within the spherical region is thus

$$\dot{E}_{\rm thm} = \eta_{\rm thm} \dot{M}_{\rm BH} c^2. \tag{12}$$

### 2.2.3. Super-Eddington feedback

Finally, in the case when $f_{\rm Edd} > 1$, energy is injected as both thermal and kinetic energy as shown in the schematic representation in Fig. 2. As AGN feedback happens on scales below $\Delta x$, a small region of injection must be chosen. Dubois et al. (2012a) show that large jet sizes $r_{\rm jet}$ lead to BH masses unrealistic when compared to observations. In addition, they find that increasing $r_{\rm thm}$ would effectively lead to a weaker impact on self-regulating the growth of BHs, thus concluding that the size of injection must be close to $\Delta x$. However, to prevent having numerical issues, we cannot use a single cell for feedback injection and to ensure that both feedback forms are well-separated, we set $r_{\rm jet} = 2\Delta x$ and $r_{\rm thm} = 3\Delta x$.

All cells that are in a cylinder of radius $r_{\rm jet}$ will only contain kinetic feedback. In all other cells within $3\Delta x$ of the BH, thermal feedback will be deposited. Since the energy deposition is different for both forms, we do not allow for thermal feedback (which would impact the temperature of the gas) to be injected in the region where kinetic feedback is also acting as the kinetic feedback already changes the temperature of the gas indirectly.

Based on Sądowski et al. (2016), the total AGN feedback injected in the super-Eddington regime reads as follows:

$$\dot{E}_{\rm sEdd} = \dot{E}_{\rm jet} + \dot{E}_{\rm thm} = (\eta_{\rm jet} + 0.5\eta_{\rm thm})\dot{M}_{\rm BH} c^2. \tag{13}$$

## 3. Setting up the halo

### 3.1. Numerical initial conditions

Sustaining super-Eddington accretion onto the BH requires strong gas inflows, which are more likely to occur in gas-rich high-redshift galaxies. The gas distribution in a low-mass isolated halo (for instance a $10^9 M_\odot$ halo) at high redshift is quickly destroyed by supernova (SN) feedback, with the result that star formation is shut off and that there is not sufficient cold and dense gas available for (super-Eddington) accretion on the BH (Dubois et al. 2015). In a more massive halo, such as one that has $10^{11} M_\odot$, the evolution is less violent, since the deeper potential well limits the impact from SN. To take advantage of this, we set up our initial conditions to represent an isolated DM halo of mass $10^{11} M_\odot$ at redshift $z = 4$.

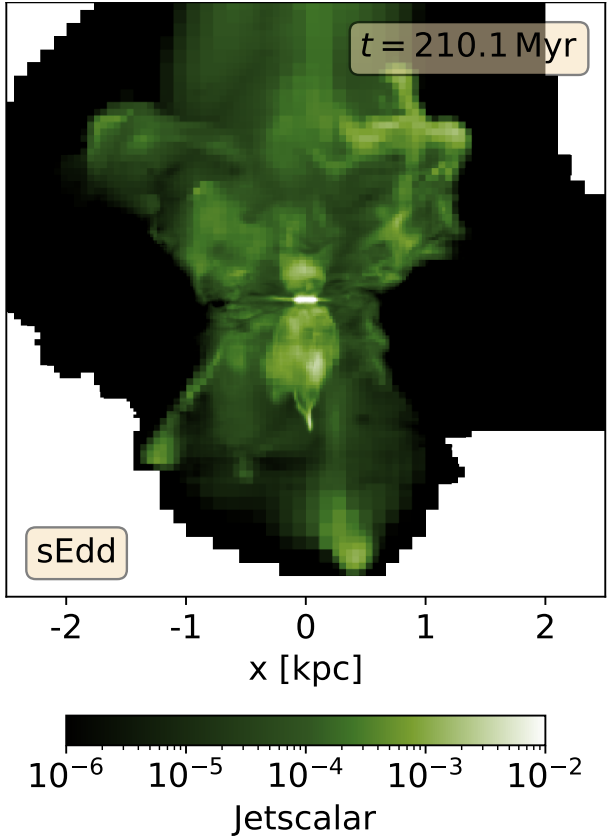

**Fig. 3.** Passive scalar used to follow the super-Eddington jet outflows at $t = 210$ Myr for the 'sEdd' simulation.

The halo is modelled using a fixed DM Navarro-Frenk-White profile (NFW, Navarro et al. 1997), centred on the box. Gas is initialised following the same density profile as the DM, with a gas fraction of $f_{gas} = 0.15$. The halo is given an initial spin of 0.02 along the $z$-axis (Bullock et al. 2001) and turbulence is injected into the gas up to 20% of the local sound speed[2]. Virial radius $R_{200}$ and concentration parameter $c_{200}$ are computed using the redshift-dependent relations from Dutton & Macciò (2014). Gas is initialised in hydrostatic equilibrium and allowed to cool from there as the simulation progresses. The average metallicity of the gas is set to $Z_{avg} = 0.3 Z_\odot$, with $Z_\odot = 0.02$ and metal-enriched gas can cool down to 10 K, using Sutherland & Dopita (1993) cooling rates down to $10^4$ K and those from Rosen & Bregman (1995) for temperatures below.

The simulations are performed in a box of size 113 kpc with a root grid of $128^3$, and then adaptively refined to a maximum resolution of $\Delta x_{max} = 12$ pc (corresponding to level 13) using four refinement criteria. A quasi-Lagrangian criterion forces the refinement of the mesh if cells which have a mass greater than $8 \times 10^5 M_\odot$. In addition to that, a Jeans length based criterion is applied: a cell is refined such that its size is smaller than the local Jeans length. Thirdly, when a BH sink particle is added, the region is automatically refined to the maximum level of refinement around it up to a spherical radius of $4\Delta x_{max}$. Finally, to refine the regions of interest affected by the jets in super-Eddington AGN feedback, we add a passive scalar variable in the BH super-Eddington kinetic outflows (following Beckmann et al. 2019), using a density $\rho_{scalar}$ equal to the gas density $\rho_{gas}$ in the region of injection. We allow for refinement if the scalar has a value above 0.01 and a variation from one cell to another greater than 0.05. This scalar follows the hydrodynamics of the gas and is able to track regions affected by the super-Eddington jets. It decays exponentially, with a decay time of $t_{decay} = 3$ Myr, making sure recent super-Eddington jet episodes are refined, while avoiding refining a large fraction of the box. This value of $t_{decay}$ was chosen to balance computational cost with the need to refine as much of the jet structure as reasonably possible. An example of the jet scalar distribution given in Fig. 3 for our fiducial simulation ('sEdd') that we will introduce later.

---

[2] This turbulence is needed to break the spherical symmetry of the initial conditions.

The BH accretion and feedback properties for all accretion regimes are described in Sects. 2.1 and 2.2 respectively. In order to help the BH particle stay bound with the galaxy, particle drag force is included (Pfister et al. 2019). We avoid applying drag force from the gas onto the BH (Dubois et al. 2013), as it sometimes causes unwanted behaviours, such as the BH to follow its own outflows, which could lead to it being ejected from the galaxy.

Stars form in cells above a gas density threshold of $n_{SF}$ following the local star formation rate density from a Schmidt law $\dot\rho_s = \epsilon_s \rho / t_{ff}$, where $t_{ff}$ is the local gas free-fall time and $\epsilon_s$ is the star formation efficiency. We adopt a gravo-turbulent star formation efficiency that depends on the gravitational binding energy of the cloud and on the turbulent Mach number (e.g. Federrath & Klessen 2012), adopting the relations and parameters as in Dubois et al. (2021). For a spatial resolution of 12 pc, the stellar mass resolution is $M_* = 10^4 M_\odot$ (therefore star formation occurs in regions with gas densities above 181 H cm$^{-3}$).

Following the energy and mechanical injection from Kimm et al. (2015), SN feedback is modelled as a total specific energy release between 3 and 50 Myr of $2 \times 10^{49}$ erg $M_\odot^{-1}$, with a semi-continuous energy injection by individual SN explosions of $10^{51}$ erg at a time with realistic time delays Kimm et al. (2015). We set the metal yield at 10% for the SN feedback. The parameters to set up the cooling isolated halo at $z = 4$ are summarized in Table 1.

### 3.2. Galaxy formation

When searching for a set of initial conditions that would allow for super-Eddington accretion onto the BH, we found that efficient star formation quickly suppressed gas inflows into the centre of the galaxy, which suppressed accretion rates onto the BH. The collapse of the halo triggers star formation in its centre, which has densities reaching $10^3$ H cm$^{-3}$. The resulting SN explosions are strong enough to heat up the gas over several kpc, eventually breaking the integrity of the halo. Since we are looking for initial conditions that allow for super-Eddington accretion, this initial SN burst, which is a consequence of having an isolated halo rather than a fully cosmological simulation, has to be circumvented. There are various options generally used in isolated simulations, such as starting with an artificially low star formation or SN efficiency and increase it to standard values over time, or alternatively do something similar with cooling. In this paper we opted to suppress the SN feedback by changing the

**Table 1.** Initial conditions for the cooling isolated halo.

| $M_{\mathrm{halo}}$ (in $M_\odot$) | $c_{200}$ | $R_{200}$ (in kpc) | $T_{200}$ (in K) | $Z_{\mathrm{avg}}$ (in $Z_\odot$) | $n_{\mathrm{SF}}$ (in H cm$^{-3}$) | $M_*$ (in $M_\odot$) | $\epsilon_{\mathrm{SN}}$ | $\Delta x_{\mathrm{max}}$ (in pc) |
|---|---|---|---|---|---|---|---|---|
| $10^{11}$ | 3.4 | 42.82 | $3.6\times 10^5$ | 0.3 | 10 | $10^4$ | 0.2 | 12 |

**Notes.** From left to right: halo mass ($M_{\mathrm{halo}}$), NFW concentration ($c_{200}$), virial radius ($R_{200}$), virial temperature ($T_{200}$), average metallicity ($Z_{\mathrm{avg}}$), density threshold for star formation ($n_{\mathrm{SF}}$), stellar mass resolution ($M_*$), SN feedback efficiency ($\epsilon_{\mathrm{SN}}$) and smallest cell size ($\Delta x_{\mathrm{max}}$).

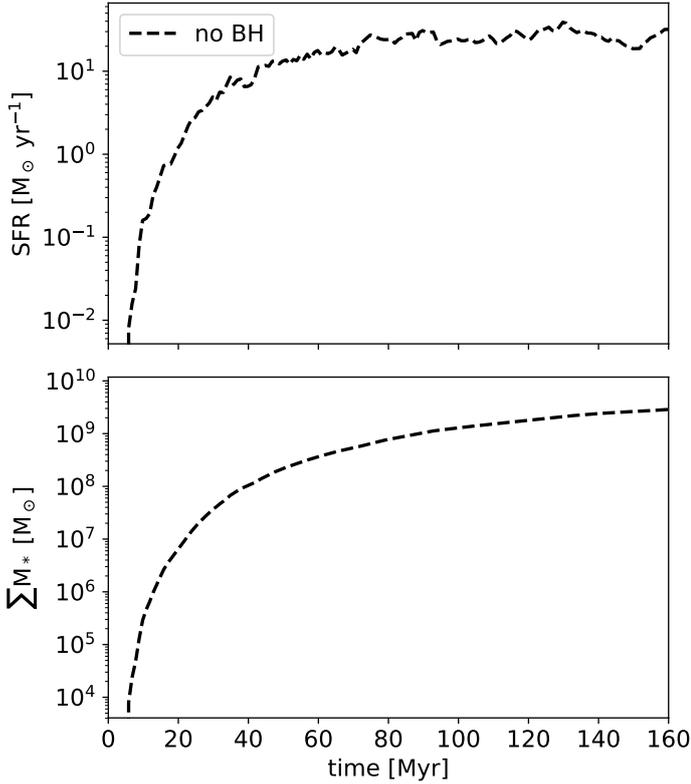

**Fig. 4.** Evolution of the SFR (*top*) and total stellar mass (*bottom*). As the halo cools, stars start to appear as conditions for their formation are met. Over the course of 100 Myr, the SFR steadily stays above 10 $M_\odot$ yr$^{-1}$, allowing the galaxy to reach $M_* = 10^9\,M_\odot$ within 160 Myr.

hydro solver to a more diffusive one when forming the galaxy. This eliminates the necessity of choosing the functional form of how to increase star formation, SN efficiency or cooling.

For this reason, we use a Lax-Friedrichs "llf" approximate Riemann solver for the early (BH-free phase) of the simulation, which allows for copious star formation as the halo collapses, without having SN feedback unbinding the gas. For the part of the simulation relevant for this experiment, i.e., when a BH and its feedback are included, we switch to the Harten-Lax-van Leer-Contact (Toro 1999) "hllc" solver, which has a less diffusive nature and is a more accurate solver for the important part of this study.

We consider our galaxy to have reached a steady state (i.e. had enough cycles of star formation and SN feedback events) when the total stellar mass reaches the stellar-to-halo mass relationship (Moster et al. 2010): the total stellar mass for a $10^{11}\,M_\odot$ halo is about $10^8$–$10^9\,M_\odot$. As shown in Fig. 4, the expected total stellar mass is reached within 160 Myr, with a quasi constant SFR of a few 10 $M_\odot$ yr$^{-1}$, which is also the expected order of magnitude for our target redshift $z = 4$ (Salmon et al. 2015).

We find that our galaxy is dominated by a central stellar proto-bulge, with a half-mass radius averaging ~110 pc at 160 Myr. The galaxy extends up to ~1 kpc from its centre of mass, as stars continue to form in the disc. The galaxy is therefore very compact, a common trait for high-redshift galaxies (Allen et al. 2017; Shibuya et al. 2019). An edge-on (top) and face-on (bottom) projection views of the galaxy gas density, temperature and stellar mass are shown in Fig. 5.

Once a galaxy-like structure with clumps in its spiral arms starts to form and settles with a realistic stellar mass and star formation rate, a BH is added in the centre of the galaxy[3]. At this point, we switch to the more accurate "hllc" approximate Riemann solver.

In order to see the effect of the change of hydro solvers mid-way through the simulation, we investigate the evolution of the SFR for both solvers at late times in Fig. 6. "hllc" (dash-dotted green) is the simulation where we swap solver at $t = 160$ Myr, and is the one used for the remainder of the work presented here, while "llf" (dashed black) continues to use the same solver throughout for comparison. In the "hllc" case the SN outflows expand further because of the less diffusive solver used. Higher diffusivity smoothes shocks, which prevents them from propagating efficiently. In a less diffusive solver, such as "hllc", the shocks are maintained and SN bubbles continue to expand due to the pressure difference inside and outside the bubble. As expected, this more efficient SN feedback somewhat decreases the SFR. This effect is notable but does not significantly impact the long-term evolution of the galaxy as the SFR is of the same order of magnitude in both cases. We therefore conclude that transitioning from an "llf" to an "hllc" solver does not produce a strong discontinuity in the evolution history of our galaxy. Changing solvers therefore allows us to have an isolated galaxy that can meet the conditions for super-Eddington accretion, while still using the most appropriate and accurate hydro solver for the part of the simulation concerned with BH accretion and feedback.

### 3.3. Adding the BH

The next step is to choose the properties of the BH to be inserted in the simulation. For galaxies with stellar mass in the range $10^8$–$10^{10}\,M_\odot$, BH masses of $10^4$ to $10^7\,M_\odot$ are possible, based on observations in the local Universe (Reines & Volonteri 2015). Since we need to have a setup that allows for super-Eddington accretion, and the Eddington limit is proportional to $M_{\mathrm{BH}}$ while the BHL accretion is proportional to $M_{\mathrm{BH}}^2$, more massive BHs are favoured.

Before testing the effect of the self-consistent AGN feedback on the BH growth, we first want to test whether the galaxy setup is favorable for BH growth, and if it can produce a sus-tained super-Eddington accretion rate in the absence of its own feedback. Therefore, in Fig. 7, we estimate the accretion properties for a potential $M_{BH} = 10^6 \, M_\odot$ BH at different points of time during the evolution of our galaxy. This is done in post-processing, and, hence, it does not include the feedback from the BH. At each point in time, we assume that the BH is located at the center of mass of the galaxy, and has a velocity equal to the averaged velocity of the gas around it (see Sect. 2.1). Having the same velocity as the gas cells around the BH maximises the BHL accretion rate $\dot{M}_{BHL}$ (blue triangle). We compare this value to the available mass $\dot{M}_{floor}$ within the accretion region of the BH (green filled circle) and to the Eddington limit $\dot{M}_{Edd}$ (dotted line). We generally find an ample gas supply to sustain BHL accretion, as $\dot{M}_{BHL}$ is generally below $\dot{M}_{floor}$. The environment is also usually favorable for super-Eddington accretion, as accretion rates on average exceed $10 \dot{M}_{Edd}$. By contrast, a similar analysis for a $10^5 \, M_\odot$ BH (not shown) leads to accretion rates that are on average below the Eddington limit (and 100 times below the blue triangles in Fig. 7), while a significantly higher BH mass of around $10^7 \, M_\odot$ would lead to accretion being limited by the local gas supply ($\dot{M}_{floor} < \dot{M}_{BHL}$, not shown).

---
[3] The centre of the proto-bulge remains close to the gas' centre of mass and to the DM centre of mass, centered on the box, throughout the simulation.

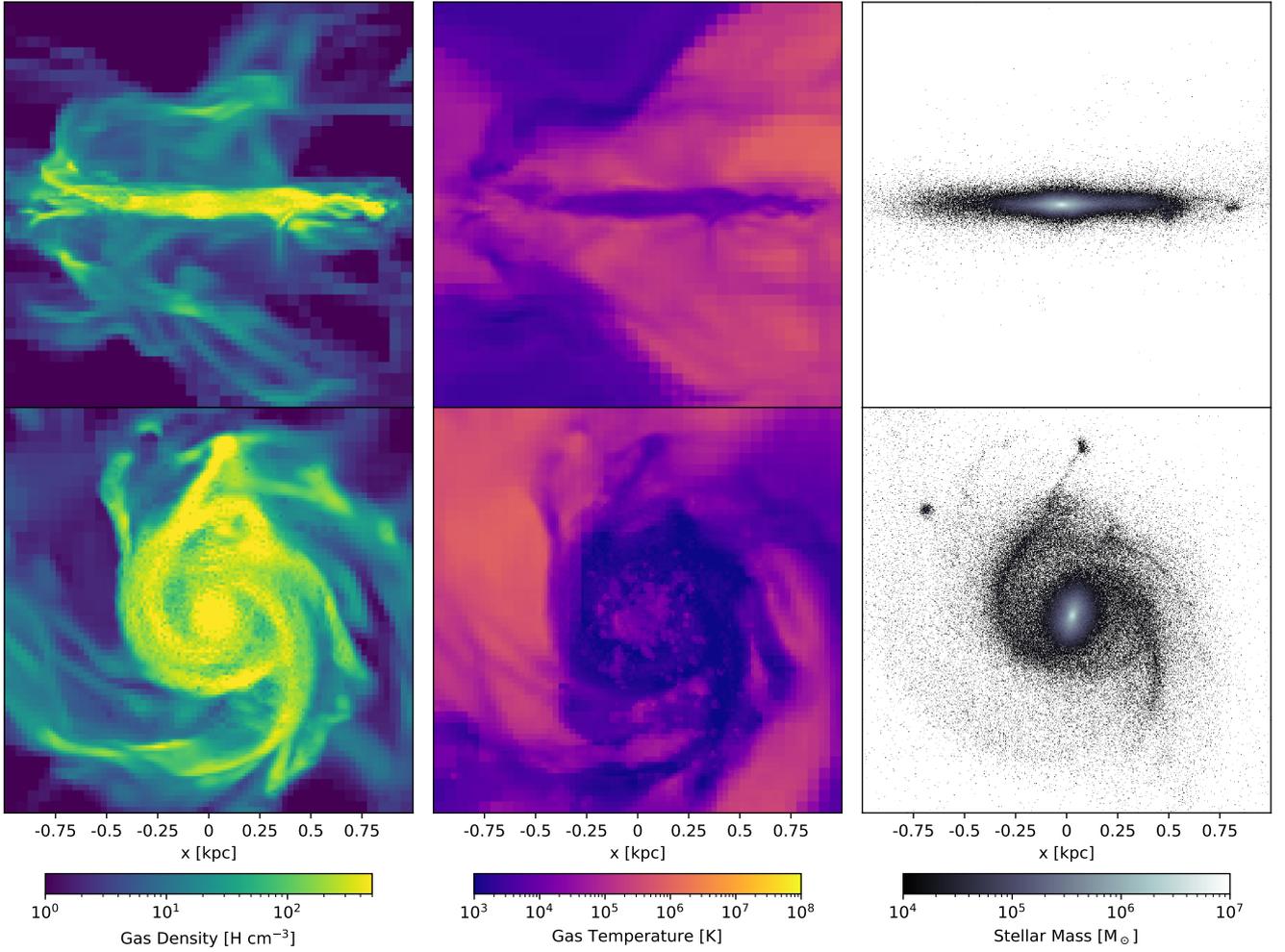

**Fig. 5.** $1 \times 1 \times 1$ kpc edge-on (*top*) and face-on (*bottom*) projections of the galaxy at the centre of the DM halo at $t = 160$ Myr. The first column shows the gas density, the second column shows the gas temperature and the third column shows the stellar mass.

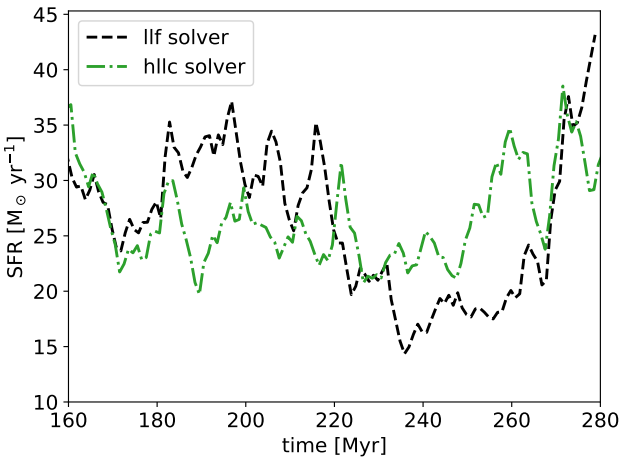

**Fig. 6.** Comparison of the evolution of the SFR between "llf" (dashed black) and "hllc" (dash-dotted green) runs. Changing the solver to one that is less diffusive does not change the SFR drastically. The SN explosions do not disrupt the environment as effectively since the galaxy is already formed.

We therefore choose to add a $M_{BH} = 10^6 \, M_\odot$ BH to our simulations and follow the same prescription described previously regarding its initial position and velocity. The normalized spin of the BH is fixed along the $z$-axis, parallel to that of the halo, and its value is set to $a = 0.7$ (which gives the canonical radiative efficiency of $\epsilon_r = 0.1$ for the standard thin accretion disc). We chose to not evolve the spin magnitude and direction in these sets of simulations presented here to highlight only the effects of super-Eddington accretion. Studying the impact of a fully self-consistent spin evolution of the BH as in Dubois et al. (2014) will be left to future work.

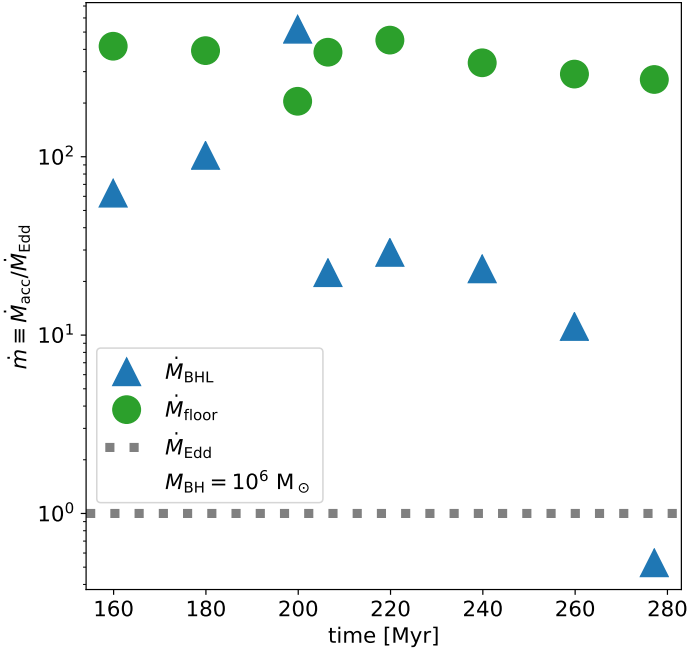

**Fig. 7.** Evolution of the normalized accretion rate $\dot{m} \equiv \dot{M}_{\rm acc}/\dot{M}_{\rm Edd}$ for the isolated $10^{11}\,M_\odot$ halo, assuming that a $M_{\rm BH} = 10^6\,M_\odot$ were present. The BHL and floor accretion rates are shown in blue and green respectively and the Eddington limit by the dotted line at 1. There is enough mass available for accretion thanks to the high density present in the central region at the start, which allows for both the BHL and floor accretion rates to be above the limit.

## 4. Results

In this section, we analyse the consequences of super-Eddington accretion coupled with its AGN feedback. We study the evolution of the SFR, the accretion history of the BH and whether the feedback can have a long lasting effect on scales ranging from pc to kpc that could impact BH growth and galaxy evolution. We explore the parameter space of the jet efficiency and also perform a resolution study to understand how super-Eddington performs at different levels of resolution. We gather all the simulations performed in Table 2.

### 4.1. Comparison between Eddington-limited and super-Eddington accretion

In order to understand the impact of the super-Eddington regime on BH growth and galaxy evolution, we perform four simulations with different accretion and feedback regimes. The first simulation allows the BH to accrete above the Eddington limit but does not include any AGN feedback ('sEddNF'). The purpose of this simulation is to establish an upper limit to BH growth in our galaxy now in a numerically self-consistent way as opposed to the previous post-processing approach (Sect. 3.3). The second simulation is Eddington-limited and releases AGN feedback according to its accretion rate ('EddLim'). Our fiducial simulation ('sEdd') uses the Eddington-limited accretion until ~206.4 Myr, and allows for super-Eddington accretion and feedback processes from this point onwards. This allows for a smooth transition from the galaxy without a BH to a BH with potentially very strong feedback. If we added the BH with immediate super-Eddington feedback, the surge of accretion would produce extremely strong feedback that would sterilize the BH environment. By enforcing a few cycles of Eddington-limited feedback, the gas and BH are able to stabilize and the galaxy evolution remains more continuous. Finally, a simulation with thermal-only super-Eddington feedback (i.e. without the jet contribution in the super-Eddington phase) is shown ('sEddThm'). It only differs from 'sEdd' by the injection of feedback in the super-Eddington phases, which is thermal feedback released within $3\Delta x_{\rm max}$ from the BH.

#### 4.1.1. Super-Eddington AGN feedback does not impact the SFR

In Fig. 8, we show the evolution of the SFR after the BH has been added to the simulation, for the 'EddLim', 'sEddThm' and 'sEdd' cases. In the 'sEddThm' simulation, we find that the SFR evolution is identical to the 'EddLim' simulation. Overall, adding a BH to the simulation does not strongly affect the SFR, despite the fact that the BH is able to deplete its immediate surroundings efficiently.

Over the course of ~80 Myr, we find that AGN feedback, whether Eddington-limited or not, does not significantly influence the average SFR of the galaxy as the inflow of cold gas from the cooling halo continues to drive star formation (SF). AGN feedback events are not able to counter this.

In Fig. 9, we plot edge-on projections maps of the gas density and temperature of the galaxy at $t = 220$ Myr. By inspecting visually these simulations, it is clear that most of the outflows correspond to SN explosions. The released SN power is $\simeq 2 \times 10^{49}\,\mathrm{erg}\,M_\odot^{-1}$ times the SFR (varying from 10 to $\simeq 40\,M_\odot\,\mathrm{yr}^{-1}$), i.e. $\simeq 10^{43}\,\mathrm{erg\,s^{-1}}$. They impact large scales because of the SNe occurring near the edges of the galactic disc. It is easier to warm up the cold and rarefied gas found in the outer galactic regions than ejecting gas from the galaxy centre, which is very dense. The collapse of the halo is not halted despite the SN explosions expanding away from the galaxy. We find that overall, the inflow rates are larger than the outflow rates as the SN-driven outflows are hot and light and therefore do not carry away a lot of mass (but see how cosmic rays accelerated in SNe can appreciably modify the wind thermodynamics, e.g. Girichidis et al. 2018; Dashyan & Dubois 2020). The shallow potential of the $M_{\rm halo} = 10^{11}\,M_\odot$ halo allows the SN-driven winds to efficiently suppress SF (e.g. Springel & Hernquist 2003; Dubois & Teyssier 2008; Scannapieco et al. 2008; Agertz et al. 2013; Hopkins et al. 2014). Over time, they cool down and fall back onto the galaxy, thus contributing to the inflows.

Besides SN feedback, a minor contribution to regulating SF comes from AGN feedback. As can be seen from Fig. 9, in the 'EddLim' run, no outflows coming from the AGN are visible at kpc scales, as they are not powerful enough to disrupt the dense gas in the galactic centre. This is not only true at the snapshot shown here but throughout the 'EddLim' simulation. By contrast, the visible super-Eddington kinetic bipolar outflows in the 'sEdd' simulation are able to reach kpc extents, with much of the energy deposited far from the galactic centre. They do not affect star-forming regions directly (Dubois et al. 2013; Gabor & Bournaud 2014) and therefore have a smaller impact on SF than SNe, as AGN jets are more spatially concentrated and do not spread out as far as SN bubbles. In the simulations presented here, the jet is always roughly perpendicular to the galactic plane, meaning gas outside of the polar region will not

**Table 2.** Properties of the suite of simulations performed.

| Name | $M_{BH}$ (in $M_\odot$) | $a$ | Accretion above Eddington | AGN Feedback above Eddington | $f_{MAD}$ | $\beta_{jet}$ | $\eta_{jet}$ | $\eta_{thm}$ | $\Delta x_{max}$ (in pc) |
|---|---|---|---|---|---|---|---|---|---|
| EddLim | $10^6$ | 0.7 | ✗ | ✗ | 0.5 | 32 | 0.16 | 0.015 | 12 |
| sEdd (fiducial) | $10^6$ | 0.7 | ✓ | Kinetic+Thermal | 0.5 | 32 | 0.16 | 0.015 | 12 |
| sEddNF | $10^6$ | 0.7 | ✓ | ✗ | ✗ | ✗ | ✗ | ✗ | 12 |
| sEddThm | $10^6$ | 0.7 | ✓ | Thermal | 0.5 | 32 | 0.16 | 0.015 | 12 |
| sEdd_0.05 | $10^6$ | 0.7 | ✓ | Kinetic+Thermal | 0.05 | 0.32 | 0.0016 | 0.015 | 12 |
| sEdd_0.1 | $10^6$ | 0.7 | ✓ | Kinetic+Thermal | 0.1 | 1.28 | 0.006 | 0.015 | 12 |
| sEdd_0.25 | $10^6$ | 0.7 | ✓ | Kinetic+Thermal | 0.25 | 8 | 0.04 | 0.015 | 12 |
| HR | $10^6$ | 0.7 | ✓ | Kinetic+Thermal | 0.5 | 32 | 0.16 | 0.015 | 6 |
| LR | $10^6$ | 0.7 | ✓ | Kinetic+Thermal | 0.5 | 32 | 0.16 | 0.015 | 25 |

**Notes.** From left to right: BH mass ($M_{BH}$); BH spin ($a$); if the super-Eddington regime is allowed; the form of feedback in the super-Eddington regime; MADness fraction of the disc ($f_{MAD}$); mass-loading factor ($\beta_{jet}$); jet feedback efficiency ($\eta_{jet}$); thermal feedback efficiency ($\eta_{thm}$); and smallest cell size ($\Delta x_{max}$). All simulations start with a $10^6 M_\odot$ BH with a fixed spin set to $a = 0.7$.

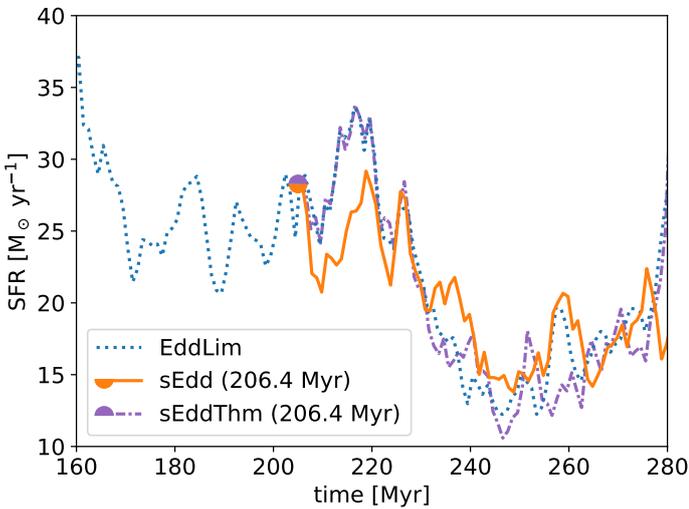

**Fig. 8.** Evolution of the SFR after 160 Myr for the simulations with accretion capped at the Eddington limit ('EddLim' – dotted blue) and above the Eddington limit with ('sEdd' – solid orange) and without ('sEddThm' – densely dash-dotted violet) kinetic feedback starting at 206.4 Myr (orange/violet filled semi-circles). The SFR does not decrease below $10 M_\odot yr^{-1}$ after the BH was added, even in the case of the super-Eddington feedback. The difference between the runs is small, showing that AGN feedback does not majorly impact the SFR in this particular setup.

be affected by the AGN over the timescale studied here. Over longer periods of time, the impact of AGN feedback will become more significant, as the AGN jet bubbles will continue to expand. Despite the fact that the jets do not contain a lot of mass, their energy is significant so they keep their regions of impact very hot because of the continuous AGN outflows in the same direction. A complete discussion of the impact of AGN feedback is given in Sect. 4.2.

In conclusion, there is little impact from the super-Eddington AGN feedback on SFR. Narrow AGN outflows cannot counter inflowing gas and most of the effect on SFR is due to SN feedback. The fact that there is no rapid quenching from AGN-driven outflows is also caused by the short timescales studied here, which do not allow us to draw firm conclusions about their long-term impact. Finally, the halo being in isolation, we cannot address the effect of feedback on the circumgalactic medium and its ability to replenish the galaxy gas supply.

### 4.1.2. Super-Eddington AGN feedback regulates BH growth

In Fig. 10, we show the mass evolution of the BH for our different accretion regimes. In the run with super-Eddington accretion but no feedback ('sEddNF' – dash-dotted green), as expected, the BH grows rapidly in mass. This BH is able to gain more than 300 times its initial mass within 100 Myr, accumulating more mass than a BH growing constantly at the Eddington limit (dashed red). Due to its very fast growth, not enough mass is available locally to accrete at the (super-Eddington) BHL rate, and the BH is limited to $\dot{M}_{floor}$. The average growth rate of this BH is almost $\sim 5 M_\odot yr^{-1}$, which corresponds to the Eddington limit of a $10^8 M_\odot$ BH. When feedback from the AGN is included, the growth of the BH is hampered. In both the 'EddLim' (dotted blue) and the fiducial 'sEdd' (solid orange) cases, we find a mass growth below $\dot{M}_{Edd}$. The 'EddLim' BH gains mass almost continuously at the limit, despite AGN feedback, up until $\sim 260$ Myr when the growth slows down, while in the 'sEdd' simulation the BH self-regulates as soon as the super-Eddington regime is turned on (at $t = 206.4$ Myr shown by the orange filled semi-circle). Overall, the BH grows more quickly in the 'EddLim' setup than in the 'sEdd' simulation, despite the fact that the BH is theoretically allowed to accrete faster in 'sEdd' than in 'EddLim'. The 'sEddThm' simulation is discussed in Sect. 4.2 below.

To better grasp what happens during the accretion phases in the simulations with AGN feedback, we compare in Fig. 11 the fraction of time spent by the BHs above a given AGN energy deposition rate $\dot{E}$ for both the 'EddLim' and 'sEdd' setups, using the same color code as in Fig. 10. This fraction of time is computed from the time when the BH is injected for the 'EddLim' simulation, or when the super-Eddington regime is turned on in the 'sEdd' and 'sEddThm' simulations, up until the end of the simulations at $t = 282$ Myr. We have added as a red patch the range of AGN feedback power for a BH (with an initial mass of $10^6 M_\odot$) constantly growing at the Eddington limit, i.e. with $\dot{E} = \eta_{thm} L_{Edd}$, for $\sim 80$ Myr. It varies between $10^{43}-10^{44}$ erg s$^{-1}$ and provides an idea of how often the BH accretes close to the limit.

As can be seen, the AGN luminosity for the 'EddLim' case remains within a narrow band around $10^{43}-10^{44}$ erg s$^{-1}$, close to the Eddington luminosity of the BH. Very few feedback events occur at luminosities below $10^{43}\,\text{erg s}^{-1}$, with the lowest luminosity at $10^{42}\,\text{erg s}^{-1}$. This means that the AGN feedback events are not strong enough to have a significant impact on the surroundings of the BH, and self-regulation is only reached towards the end of the simulation as shown in Fig. 10.

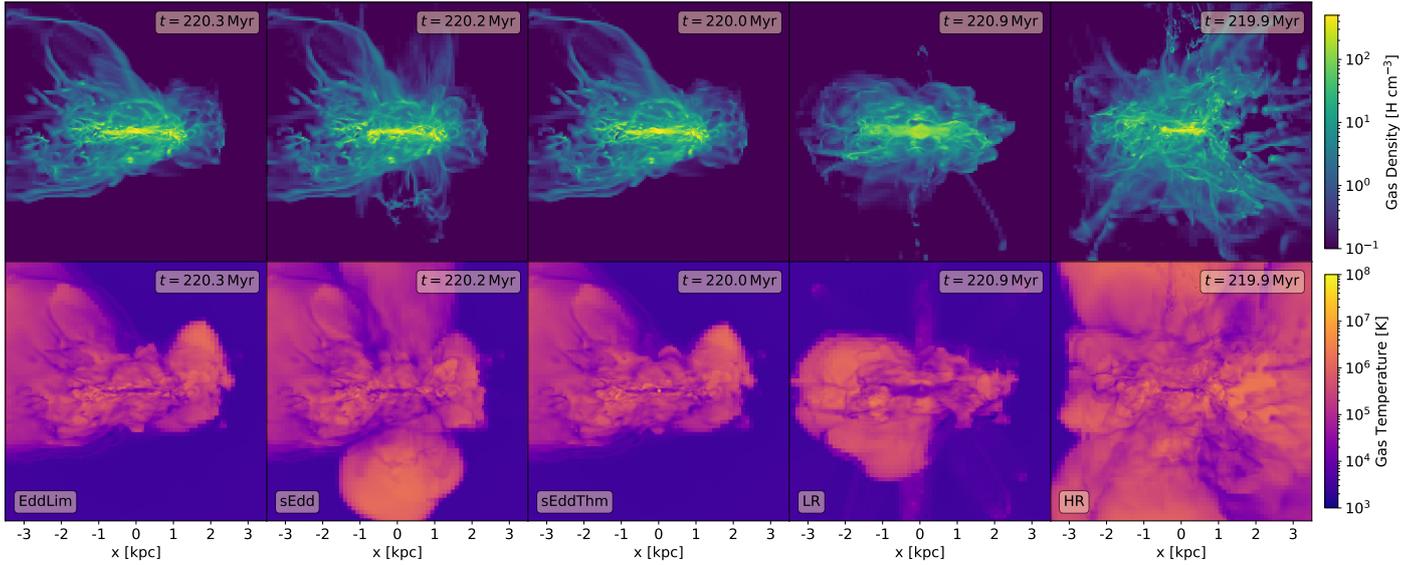

**Fig. 9.** *Left to right*: edge-on projection maps of the 'EddLim', 'sEdd', 'sEddThm', 'LR' and 'HR' simulations at $t = 220$ Myr. Gas density and temperature are respectively shown in the first and second row. SNe as well as super-Eddington kinetic feedback are visible on large scales.

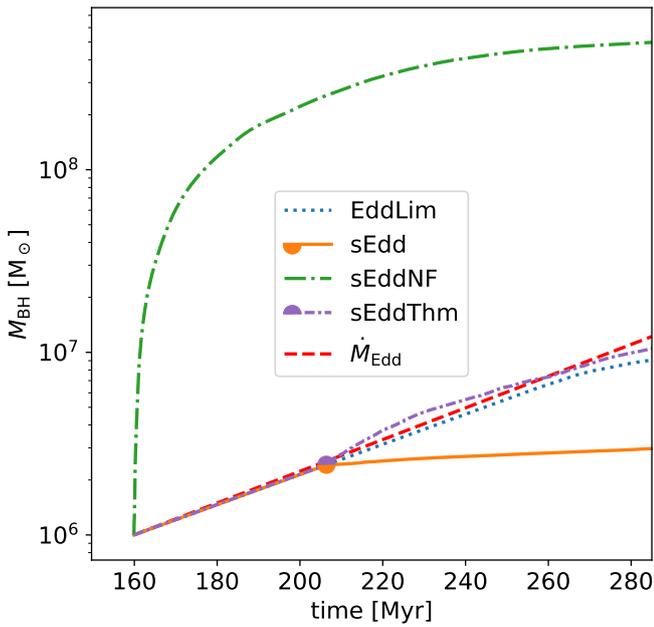

**Fig. 10.** Evolution of the BH mass for the 'EddLim' (dotted blue), 'sEdd' (solid orange), 'sEddThm' (densely dash-dotted violet) and 'sEddNF' simulations (dash-dotted green). Both 'sEdd' and 'sEddThm' simulations start at 206.4 Myr (colored filled semi-circles). In dashed red is also added the expected evolution of a $10^6\,M_\odot$ BH if it accreted constantly at the Eddington limit. It is clear that super-Eddington accretion, coupled with feedback does not help the BH to grow. We note that without jets (i.e. 'sEddThm' run), BHs with super-Eddington accretion are less affected by their feedback and are able to grow above the limit for at least 60 Myr. In fact, the BH reaches a higher mass than the one in 'EddLim' (see Sect. 4.2 for more details).

When allowing for super-Eddington feedback ('sEdd'), some strong feedback events occur ($\dot{E} > 10^{44}\,\text{erg s}^{-1}$). These powerful episodes have a wide range of luminosities, varying from $10^{44}$ to $10^{46}\,\text{erg s}^{-1}$. As a result of these brief super-Eddington phases, the AGN spends a significant fraction of the remaining time in a low luminosity state ($10^{36} < \dot{E}/(\text{erg s}^{-1}) < 10^{43}$). A small number of high luminosity events with strong feedback have overall a stronger effect than a high number of intermediate luminosity events with weaker Eddington-limited feedback. This shows that super-Eddington feedback more efficiently self-regulates BH growth than sub-Eddington feedback.

To further investigate this phenomenon, we show in Fig. 12 the evolution of the AGN Eddington ratio during a short period of 8 Myr, representing a succession of typical super-Eddington episodes. We also added the 'EddLim' run for comparison. In the Eddington-limited run, although the released power keeps increasing steadily as the BH grows, the BH emits almost constantly at the Eddington luminosity, since feedback does not produce a strong effect on the BH environment. In the 'sEdd' case the Eddington ratio fluctuates rapidly, by up to five orders of magnitude. Every single super-Eddington event is followed by a sharp and instantaneous drop down to very low luminosities. At our resolution and with our set up, the slow rises and sharp drops visible in the zoomed inset are characteristic of the bursts of super-Eddington activity. Studies in RHD with more idealized set-ups but that resolve $r_\text{BHL}$ and the associated timescales (e.g. Park & Ricotti 2011; Park et al. 2020) instead find sharp rises and slow drops, when the Strömgren radius is resolved. Takeo et al. (2020) find a similar oscillatory behaviour when momentum and radiation are injected in an axisymmetric two-dimensional simulation. In our simulations the unresolved $r_\text{BHL}$ and sound-crossing time at this scale, as well as the direct deposition of feedback energy lead to our inability to resolve a possible similar oscillatory behaviour. We cannot therefore assess if slow rises and sharp drops in the accretion rate occur also in conditions characterized by a turbulent and non-smooth gas distribution in three dimensions and in the presence of relativistic jets. Nevertheless, this confirms that the balance between the high power in the short-lived super-Eddington episodes and the lower power in the longer-lived sub-Eddington phases is such that globally this type of AGN feedback is more effective at curbing BH accretion even with overall smaller energy injection.

We also notice that the succession of these super-Eddington episodes is short, with the intervals between them varying from 1 Myr down to 20 kyr. The spread in these values reflects in part the complex and multiphase structure of the interstellar medium in the galaxy, and in part the strength of the AGN feedback itself: the more powerful the super-Eddington event, the longer the time before the next super-Eddington episode. Furthermore, during a super-Eddington phase of a few 0.1 Myr (as shown in the inset), there are multiple bursts of super-Eddington accretion interleaved by ∼10 kyr that are growing in strength until the last super-Eddington event is strong enough to quench accretion for a few 0.1 Myr. The multiple successive bursts occur when feedback is unable to evacuate the accretion region (a sphere of radius of $4\Delta x$ with weight given by Eq. (3)). In this case the periodicity is shorter than the free-fall time at the level of one resolution element, therefore it is arguably caused by gas mixing within the accretion region. After a powerful episode where the full accretion radius is evacuated, the time to the next super-Eddington episode is of the order of the free-fall time from the edge of the evacuated region, which has an extent of several tens to hundreds of parsecs.

In conclusion, we find that the super-Eddington AGN feedback in this idealised setup does not allow the BH to grow efficiently, because each episode of super-Eddington accretion is followed by a period of very sub-Eddington accretion. This is a consequence not only of the amount of energy injected, but also of the form of injection, as we will demonstrate in Sect. 4.2.

### 4.2. Importance of the kinetic AGN feedback in the super-Eddington regime: impact from pc to kpc scales

#### 4.2.1. Thermal/Kinetic AGN feedback: impact on the BH growth

Observationally, super-Eddington AGN feedback does not always produce a strong kinetic bipolar radio-like outflow, as for instance shown by the super-Eddington but radio-quiet AGN sample presented in Du et al. (2014). The lack of jet in such objects can be caused by zero or very low spins, or by unfavourable magnetic field configurations (e.g. Beckwith et al. 2008; McKinney et al. 2012). To explore how thermal-only super-Eddington feedback impacts BH growth, we analyse the 'sEddThm' run in comparison with our fiducial 'sEdd' run. 'sEdd' in fact has a kinetic feedback approximately 20 times more powerful than the thermal injection of energy because of the different feedback efficiencies (see Sect. 2.2 and Table 2).

Starting at the same time for both simulations, the BH producing only thermal feedback ('sEddThm' – densely dash-dotted violet) grows much faster than our fiducial run ('sEdd' – solid orange), as shown in Fig. 10. In the 'sEddThm' run, the mass of the BH immediately increases above the Eddington limit and is able to triple its mass from the first super-Eddington phase, since the inflow of cold gas is not counteracted as effectively as in the presence of kinetic feedback. We also find that thanks to this initial super-Eddington mass growth, the BH is ∼15% more massive than the one in the 'EddLim' run after ∼80 Myr of growth. However, BH mass growth in 'sEddThm' is not able to keep pace with the theoretical growth at the Eddington limit, whose mass gain exceeds that in 'sEddThm' after ∼60 Myr. The limiting factor that reduces the accretion rate is not the amount of available gas but its temperature.

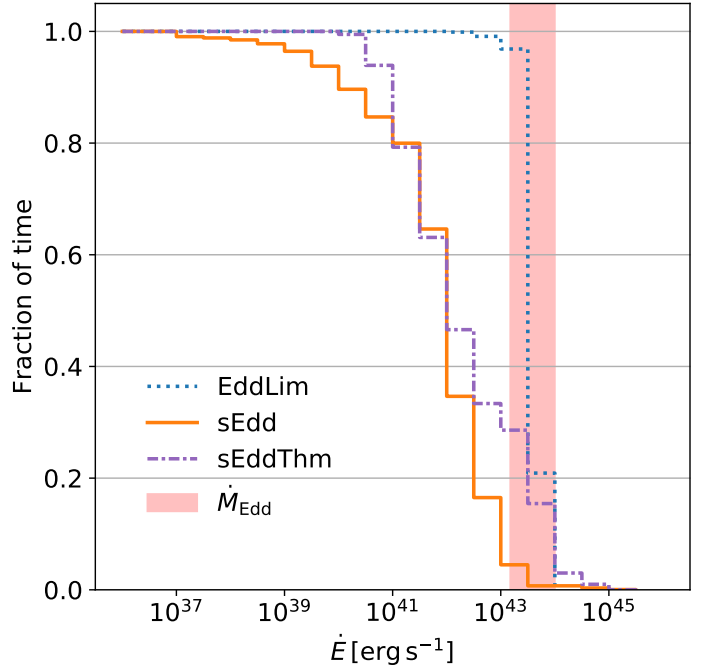

**Fig. 11.** Fraction of time spent by the BHs above a given AGN energy injection rate $\dot{E}$ for the 'EddLim' (dotted blue), 'sEdd' (solid orange) and 'sEddThm' (densely dash-dotted violet) simulations. A red patch $\dot{M}_{\rm Edd}$ shows the AGN feedback range ($\sim 10^{43}-10^{44}$ erg s$^{-1}$) of a BH between $M_{\rm BH} = 10^6-10^7\,M_\odot$ at the Eddington limit. The 'EddLim' simulation has a BH constantly accreting at the limit, thanks to the ineffectiveness of the feedback. The 'sEdd' case has stronger feedback events but this results in many low luminosities feedback episodes. This regime therefore reduces the overall required AGN feedback to self-regulate the BH growth. The 'sEddThm' simulation is discussed in Sect. 4.2.

As discussed in Sect. 4.1, in the 'EddLim' simulation, the BH is for the most part in quasar mode, which also has purely thermal energy injection, and the impact of feedback in 'EddLim' is very limited over long timescales. In 'sEddThm' the impact of feedback is stronger because the injected peak power is greater due to larger reached accretion rates in the super-Eddington regime. The feedback power reaches $\sim 10^{45}$ erg s$^{-1}$ during the super-Eddington phases, as shown in Fig. 13, over an order of magnitude above the peak luminosities in 'EddLim' and of the same order of magnitude as 'sEdd'.

There are however noticeable differences between the duty cycles of the 'sEdd' and 'sEddThm' runs, such as in the frequency at which super-Eddington episodes occur. Intervals between two episodes of the 'sEddThm' run average 40 kyr, sometimes with consecutive super-Eddington phases. This is much more frequent than in the fiducial setup, indicating a lower impact of feedback on the environment. Combining these results with Fig. 11, we find that the 'sEddThm' AGN feedback appears less destructive than the jetted case. This can been seen by the fact that the 'sEddThm' simulation shows more instances of high luminosities ($\dot{E} > 10^{44}$ erg s$^{-1}$) and a higher minimum luminosity ($\dot{E} = 10^{41}$ erg s$^{-1}$). As a result, with only thermal feedback, the AGN is both more likely to be in a super-Eddington state and has a higher luminosity on average, even discounting the super-Eddington events. These events nevertheless, are less effective in impacting their environment compared to the 'sEdd' case. As a result, the BH in the 'sEddThm' run grows more efficiently above the Eddington limit.

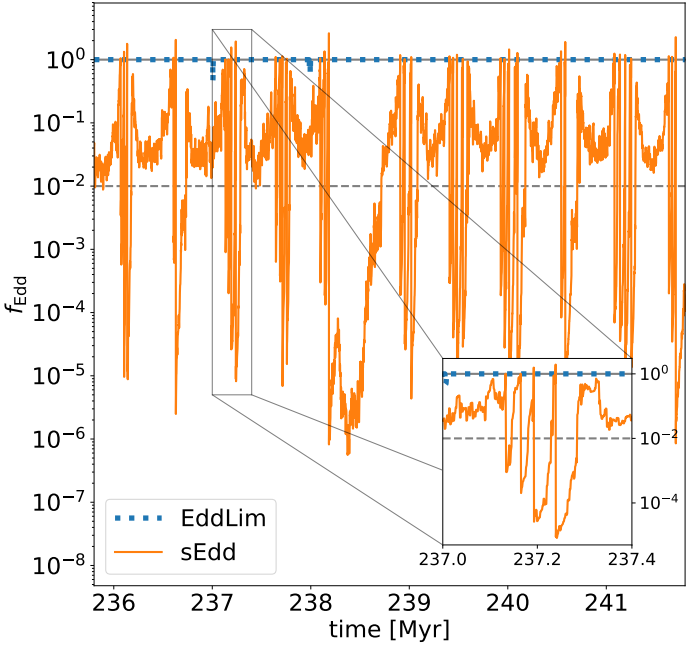

**Fig. 12.** Evolution of the AGN Eddington fraction $f_{\rm Edd}$ for the 'EddLim' (dotted blue) and the 'sEdd' (solid orange) simulations over 8 Myr. Powerful and instantaneous super-Eddington episodes lead to sharp drops in $f_{\rm Edd}$, reducing the AGN feedback energy injected overall. Solid and dashed horizontal lines indicate $f_{\rm Edd} = 1$ and 0.01 respectively, that mark the separations between the different modes of accretion: super-Eddington, quasar, and radio modes (*from high to low* rates).

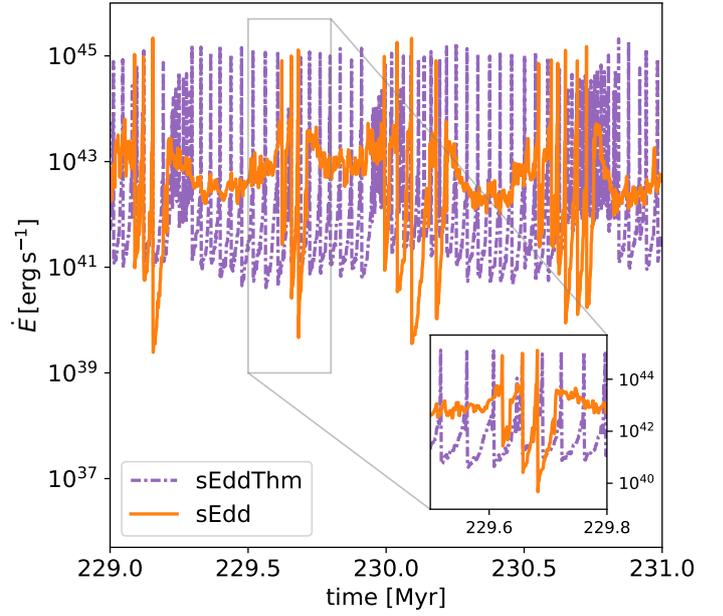

**Fig. 13.** Comparison between the energy deposition rate $\dot{E}$ from the 'sEdd' (solid orange) and 'sEddThm' (densely dash-dotted violet) simulations, over a succession of typical super-Eddington episodes. The events deposit the same amount of energy, though in different ways, but the 'sEddThm' case seems to be more frequently in a super-Eddington regime. In addition, the drops post super-Eddington events do not fall to luminosities below $10^{40}$ erg s$^{-1}$.

To emphasize this change in how many super-Eddington events occur during the simulation, we show in Fig. 14 the total mass gained by the BH for the 'sEdd' and 'sEddThm' simulations after the super-Eddington regime is turned on, as well as the fraction of time that the BH spends in each feedback mode (radio, quasar and super-Eddington). We find that in both simulations, the strongly sub-Eddington regime (radio mode) does not contribute significantly to the mass growth of the BH ($\lesssim 0.4\%$), despite the BH spending a significant amount of time in this mode, as shown in the right panel.

In the 'sEdd' run, the BH gains more than 80% of its mass when close to the Eddington luminosity (in quasar mode) and spends almost 70% of its time in this regime. Only ~15% of the total mass is accreted in episodes above the Eddington limit as the BH spends less than 1% of its time in this regime.

In the 'sEddThm' run, the behaviour is different. The BH accretes much less of its total mass in the quasar mode, as it also spends less time in this regime with a fraction of time spent ≳30%. Most of the mass is gained in the super-Eddington regime, whilst the time that the BH spends in this mode increases by a factor 10 in comparison to the 'sEdd' run: the BH is able to grow above the limit for a long period of time because of the weaker feedback. This factor of 10 comes from the combination of BH spending more time in the super-Eddington regime and breaking the limit more frequently.

In summary, if super-Eddington accretion is not accompanied by powerful jets, then the BH can grow up to significantly higher masses before reaching self-regulation. The BH in run 'sEddThm' is able to reach a higher mass than the one in 'EddLim', but eventually drops below the theoretical curve assuming constant accretion at the Eddington rate. Despite not spending more than 10% of its time above the limit, the BH is able to accrete most of its mass above the Eddington limit. This is because thermal feedback is weaker and cannot push gas efficiently and create outflows like the kinetic feedback does in the 'sEdd' run, which we will discuss further in the following sections.

### 4.2.2. Impact of the AGN feedback on the BH environs

To understand how the gas behaves around the BH and how it is impacted by the repeated super-Eddington episodes, in this section we investigate the gas properties at tens of pc scales for both 'sEdd' and 'sEddThm' simulations. The AGN feedback, whether kinetic or thermal, impacts the temperature of the gas surrounding the BH, with higher energy producing higher temperatures. Outflows are also to be expected to affect the gas density, so we show in Fig. 15 the evolution of both the gas average density $\bar{\rho}$ and temperature $\bar{T}$ around the BH, in a $4\Delta x_{\rm max} = 48$ pc radius, between 204 and 212 Myr.

Before $t = 206.4$ Myr (orange/violet filled semi-circles), the BH accretion is Eddington-limited. As discussed in Sect. 4.1.2, the AGN feedback released in this regime does not impact the gas in the injection region, as the gas there remains very dense (almost $10^3$ H cm$^{-3}$) and cold (~$10^2$ K). Looking at the temperature evolution, as soon as the BH is allowed to accrete above the Eddington limit, an instantaneous increase of >6 orders of magnitude in temperature (from ~$10^2$ K to $10^8$–$10^9$ K) is visible in both simulations. This sudden rise in temperature drastically shrinks $r_{\rm BHL}$ and is the main reason for the sharp drop[4] in accretion rate (see Fig. 12), as $\dot{M}_{\rm BHL} \propto T^{-1.5}$.

---
[4] The erratic changes in temperature are not linked to the weighted kernel used by the accretion routines, as we checked by using a fixed kernel set to $2\Delta x_{\rm max}$ at all times. The behaviour of this test case was similar.

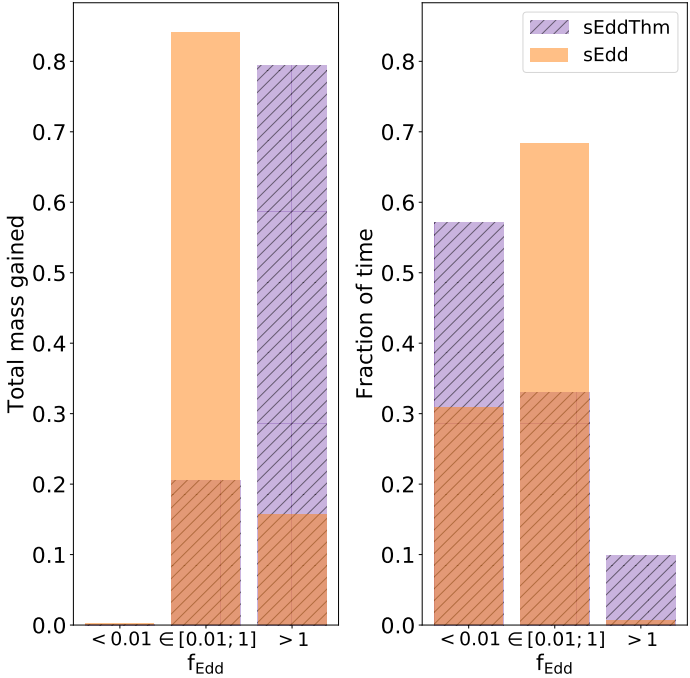

**Fig. 14.** Fractional values of the total mass gained (*left*) and fraction of time spent (*right*) by the BH in the 'sEdd' (filled orange) and 'sEddThm' (hashed violet) runs in $f_{Edd}$ bins, mirroring the three accretion/feedback regimes. Despite not spending much time in the super-Eddington regime, this mode ($f_{Edd} > 1$) still contributes significantly to increasing the mass of the BH, and dominates the mass growth in the 'sEddThm' case.

The density also varies in both simulations, but in a different way. In the fiducial 'sEdd' run, a decrease of 2 orders of magnitude in density indicates the presence of a strong outflow. The immediate increase in accretion rate suddenly drains the cells near the BH, while shortly thereafter the density drops due to the release of the powerful feedback in the super-Eddington phase. The sudden drop in density caused by this first super-Eddington outflow is clearly visible at $t = 206.8$ Myr the gas density map, shown in Fig. 16. The delayed decrease in density in comparison to the sudden change in temperature is caused by the shock propagation, as it takes more time for the outflow to propagate and escape the region.

At ~208 Myr in the 'sEdd' run, after the initial super-Eddington outflow, the density of the medium rises again to ~$10^2$ H cm$^{-3}$ and the temperature of the gas slowly decreases. This can indicate very rapid cooling or an inflow of new, colder gas onto the BH, or a combination of both. The dominant factor appears to be inflow of new gas, as to cool down gas at these densities/temperatures would require at least 0.1–1 Myr (assuming the maximum density of $10^3$ H cm$^{-3}$), which is much longer than the timescales over which changes in temperature are seen here. This inflow comes from galactic gas that has not been impacted by feedback from the BH: since the BH spin is fixed along the $z$-axis, the jet direction does not change and gas in the equatorial direction can reach the galaxy centre. The constant inflows therefore provide more material for subsequent super-Eddington episodes, as visible at $t = 209.8$ and $t = 210.8$ Myr. Furthermore, as cold and dense gas slowly reaches the BH accretion region, it triggers a succession of super-Eddington episodes (see the inset in Fig. 12) with feedback energies unable to eject gas outside of the accretion region. Mixing with infalling gas, the super-Eddington episodes grow in strength and are able to push gas outside of this region in the equatorial direction, up to a distance determined by how powerful the last super-Eddington event is (e.g. Costa et al. 2014). The results of this setup are therefore in between those derived by Regan et al. (2019) and Takeo et al. (2020): in these idealized conditions, where the galaxy presents a well-defined disc and the jet is launched perpendicularly to the disc the jet's feedback affects accretion, but inflows from the equatorial plane replenish the gas reservoir.

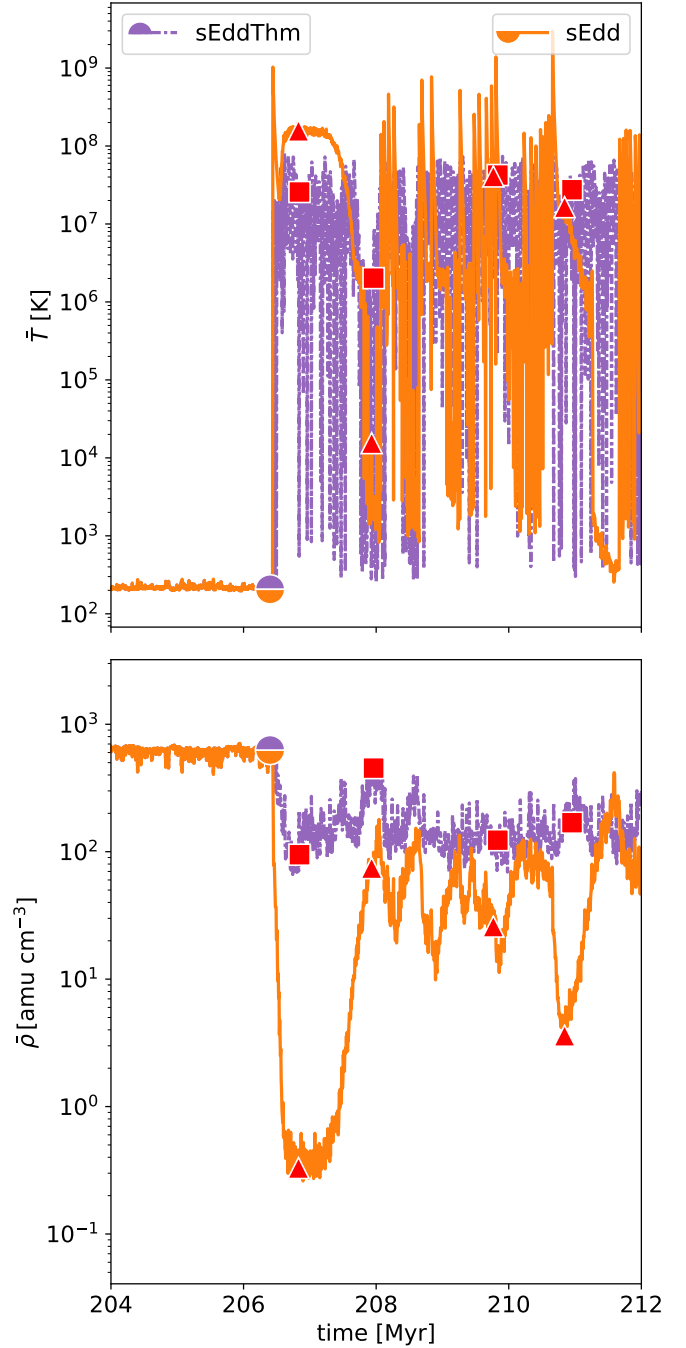

**Fig. 15.** Evolution of the average temperature (*top*) and density (*bottom*) around the BH in the accretion region for the 'sEdd' (solid orange) and 'sEddThm' (densely dash-dotted violet) simulations. The red triangles/squares correspond to the snapshots shown in Fig. 16 for the 'sEdd' and 'sEddThm' simulations respectively. As soon as strong super-Eddington events occur, a peak in temperature and a drop in gas density are observed. When kinetic feedback is involved, outflows are created by the momentum carried by the jet. But thanks to the rapid gas infall and refilling of the accretion region, the BH is able to go back in a super-Eddington phase within ≲1 Myr.

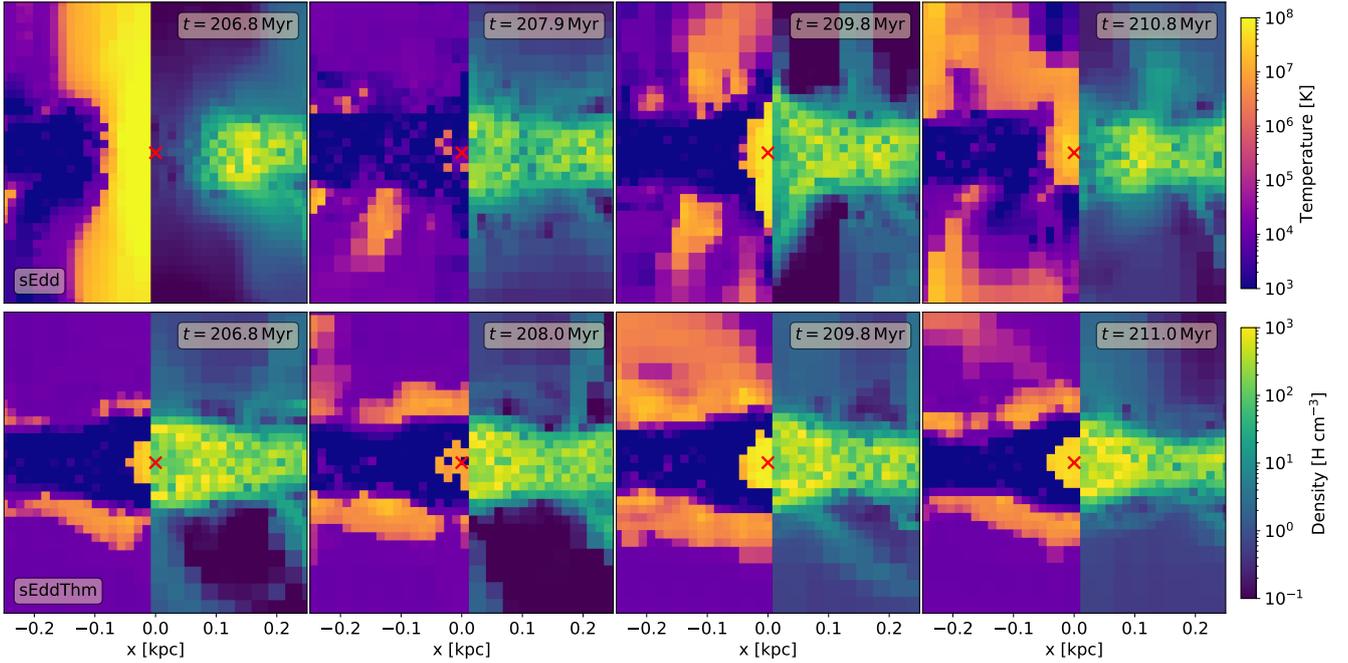

**Fig. 16.** $0.5 \times 0.5 \times 0.5$ kpc edge-on slice maps centered on the BH (red cross) of the 'sEdd' (*top*) and 'sEddThm' (*bottom*) simulations. Each panel is split in half, with gas temperature (*left*) and density (*right*). The thermal-only super-Eddington feedback creates a "bubble" unable to push any gas to larger scale, in opposition to the kinetic feedback, strong enough to escape the vicinity of the BH.

Looking closely at the 'sEdd' run in Fig. 16, we see a positive effect from the super-Eddington AGN feedback on the gas density during the bursts at $t = 206.8$ and $t = 210.8$ Myr. Density is significantly larger ($10^2$–$10^3$ cm$^{-3}$) in the ISM at the interface with the AGN outflow, which may induce a short increase in SF (Gaibler et al. 2012; Bieri et al. 2016).

On the other hand, in the 'sEddThm' simulation, despite having a BH able to accrete above the limit which reduces the density around the BH by a factor 3, the produced feedback does not provides enough momentum to the gas to create an outflow fast enough to propagate beyond the vicinity of the BH. The feedback episodes create small local "bubbles" (visible as small pockets of a few tens of pc radius, in each panel of Fig. 16 for the 'sEddThm' simulation) of hot gas (which reaches up to $10^8$ K), unable to escape the dense medium around it. They quickly vanish, as cold gas swirls in to feed the BH. This situation is a more powerful re-scaled version of the processes discussed for the 'EddLim' case.

Overall, in both simulations, the infall of cold and dense gas is able to quickly replenish the surroundings of the BH, whether gas was pushed outside of the galaxy due to super-Eddington kinetic feedback events, or simply a local small bubble of gas was heated by the thermal feedback. When more low temperature gas is available, the next super-Eddington episode ensues, resulting in a sequence of super-Eddington and highly sub-Eddington events. A parallel can be drawn between the difference in total mass gained (left panel of Fig. 14) and the density variations: in the fiducial 'sEdd' run, density drops significantly after each super-Eddington event, because gas is displaced by the outflows driven by injected momentum injected; whereas in the 'sEddThm' simulation, gas does not participate in a large-scale outflow and remains confined in the central region. In this case, the density does not vary as significantly, leading to higher accretion rates and more frequent super-Eddington episodes.

### 4.2.3. Impact of AGN feedback on galactic scales

Despite the similar injected power, thermal feedback is less effective at disrupting the regions further away from the BH. To explore this, we set up a cylinder of height = 8 kpc and radius = 1 kpc centered on the BH and perpendicular to the galactic plane. Since the visible AGN outflows (Fig. 9, 'sEdd' run) are aligned with the $z$-axis of the box, we analysed the mass-weighted velocity in this direction at different heights of the cylinder, both for the inflowing (solid) and outflowing (dashed) gas in Fig. 17. The region above/below the galaxy plane is in the positive/negative direction of the $z$-axis respectively. For each inflow/outflow, we include a "reference" case corresponding to gas before the super-Eddington episodes (measured at $t = 206.4$ Myr). After 2.5 Myr, both simulations have undergone at least one super-Eddington episode, and we show the gas velocity evolution with colored curves.

The diagram can be split into three parts. The central $\pm 0.1$ kpc of the galaxy where the BH lies corresponds to the region where feedback is injected, and has been discussed in Sect. 4.2.2. The SN feedback is not visible in this region, as the gas is very dense and the explosions are stifled. The gas remains cold, and only super-Eddington AGN feedback events can heat up this small region. The $z$-velocity of the gas in the accretion region is almost negligible, since the infalling gas coming from both above and below the disc meets in the equatorial plane.

Further away, between $\pm 1.5$ kpc, the gas is mostly impacted by the SN explosions on the edges of the disc which heat up and disrupt the colder gas surrounding the galaxy. Temperatures are higher than in the central region, as SN feedback propagates more easily in this lower density gas. The outflowing gas $z$-velocity is on average lower than the inflows as SN feedback meets with colder, infalling halo gas, which slows down the outflow and prompts them to fall back onto the galaxy. In addition, since SNe represent the dominant form of feedback in this region, we do not find significant differences regarding the outflows between the simulations. We also find that the inflowing gas $z$-velocity tends to increase in all simulations, from ±0.1 to ±1.5 kpc. There are less outflows affecting gas further away from the galaxy, and since the halo continues to collapse, the infalling $z$-velocity increases.

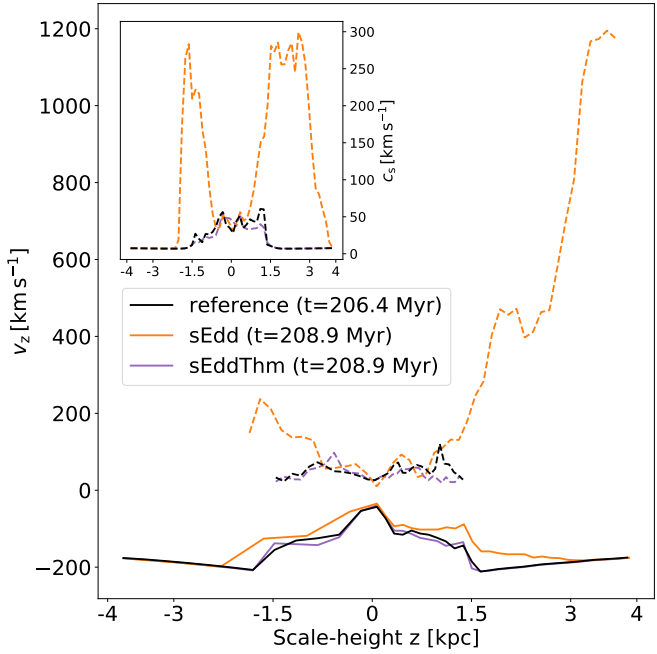

**Fig. 17.** Evolution of the average $z$-velocity $v_z$ above and below the galaxy to follow the gas outflows (dashed) and inflows (solid) for both simulations at $t = 208.9$ Myr. A "reference" curve (black) is drawn to represent the state of the gas before the first super-Eddington episode, i.e. at $t = 206.4$ Myr. Kinetic jets are clearly visible as they impact gas that is far away from the galactic disc, heating it up on their way out. On the other hand, thermal feedback does not have enough momentum to have said impact.

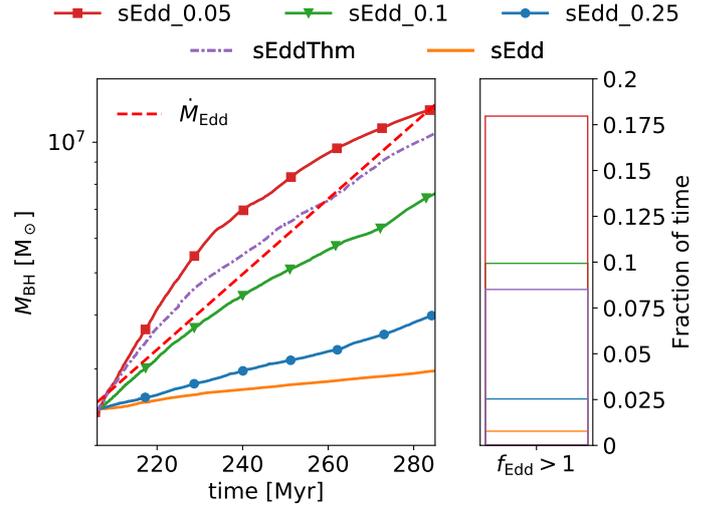

**Fig. 18.** *Left*: evolution of the BH mass for the 'sEdd_0.05' (red square), 'sEdd_0.1' (green triangles) and 'sEdd_0.25' (blue circle) from the moment super-Eddington was allowed ($t = 206.4$ Myr). For comparison are added 'sEdd' (solid orange) and 'sEddThm' (densely dash-dotted violet) as well as the Eddington limit $\dot{M}_{\rm Edd}$ (dashed red). *Right*: fraction of time spent in the super-Eddington regime for the same simulations (same colors). Weaker super-Eddington kinetic feedback affects positively the BH growth and increases the fraction of time spent in the super-Eddington regime.

Finally, at scale heights above 1.5 kpc, we have a region where gas collapses from the halo. The inflowing gas is mostly unperturbed by feedback events, as this region has no star formation, and no SN feedback is able to reach such distances. The 'sEddThm' run shows no traces of outflows at >kpc scales (see also Fig. 9, 'sEddThm' run), despite having many more super-Eddington accretion episodes, as discussed in Sect. 4.2.2. The only feedback events that are able to propagate this far in such a short period of time (2.5 Myr), are the super-Eddington driven bipolar kinetic outflows from the 'sEdd' simulation.

The initial 'sEdd' super-Eddington episode produces an outflow powerful enough to entrain dense gas from the galaxy. It has an initial velocity close to 0.1 c but encounters dense gas when it pierces through the galaxy. This creates shocks and slows the jet down considerably, to a few $10^2-10^3$ km s$^{-1}$. Because of the different gas structures encountered on the way, the AGN outflow above and below the galaxy is not symmetric. The jet below the galaxy (negative scale heights) struggles to find its way through the dense gas. After 2.5 Myr, it reaches ~1.5 kpc and almost stops, with $v_z \simeq 200$ km s$^{-1}$. It spreads out radially and mixes with the SN explosions. On the other hand, if a jet can push through this dense layer, it carves out a low-density region and creates a path for future outflows, which is what the initial outflow ejected above the galaxy does. It travels up to 4 kpc within 2.5 Myr, while keeping a steady velocity close to $10^3$ km s$^{-1}$ as it stays collimated.

In the inset, where the sound speed $c_s$ is plotted, we find that the kinetic outflow is able to heat up the region through which it passes. In comparison to the "reference", pre-super-Eddington, and 'sEddThm' curves, which are very similar and have $c_s \leq 50$ km s$^{-1}$ everywhere, the regions in 'sEdd' which were in contact with the jet have all heated up to high temperatures. Because of long cooling times, the gas in these regions does not have the chance to cool down, and subsequent super-Eddington outflows will prevent any long-term cooling. We also find that despite slowing down and spreading out, the kinetic outflow below the galaxy still efficiently heated up its surroundings to much higher temperatures than SN feedback alone.

Finally, we compare the galactic scale inflows for both simulations and find that they are not greatly affected by the super-Eddington kinetic feedback. More specifically, the timescales shown here are too short to see this effect. However, as time goes on, the jets start expanding outwards, and shutting down the gas inflows from the collapse. This happens, for the 'sEdd' simulation, ~15 Myr after the first super-Eddington jet. We do not see any impact on inflows throughout the entirety of the 'sEddThm' simulation.

In summary, galactic scales outflows are only visible if there are super-Eddington jets that are able to punch through the galactic disc. The momentum given by the super-Eddington kinetic feedback is enough to push gas to kpc scales within short periods of time, whilst staying fast (~$10^3$ km s$^{-1}$) and collimated, as long as the gas met on the way is not too dense. AGN outflows heat up the gas to very high temperatures, but because of the collimated nature of the jets, it has little impact on the gas inflows.

### 4.3. Varying the AGN feedback efficiency

Our simulations showed that super-Eddington jets are very destructive, and efficiently prevent BH growth, but that the BH is able to grow above the Eddington limit, for a long periods of time, when emitting only thermal feedback. Recently, Yao et al. (2021) detected a pc-scale jet from a super-Eddington source, hinting at the fact that super-Eddington jets are not always powerful. We explore this by changing the MADness factor $f_{MAD}$ for the jet efficiency $\eta_{jet}$. We run three other simulations, namely 'sEdd_0.25', 'sEdd_0.1' and 'sEdd_0.05' with $f_{MAD}$ = 0.25, 0.1 and 0.05 respectively (with corresponding jet efficiencies $\eta_{jet}$ of 0.04, 0.006, and 0.0016). The efficiency of the thermal feedback is not changed. As shown in Table 2, with these choices of $f_{MAD}$, we are able to investigate a range of jet properties, from weak to strong ones. For this study we compare these simulations to 'sEdd' and 'sEddThm', which are the extremes (very strong and non-existent jet respectively) of the other three cases presented here. We recall here that our fiducial run 'sEdd' has $\eta_{jet} = 0.16$.

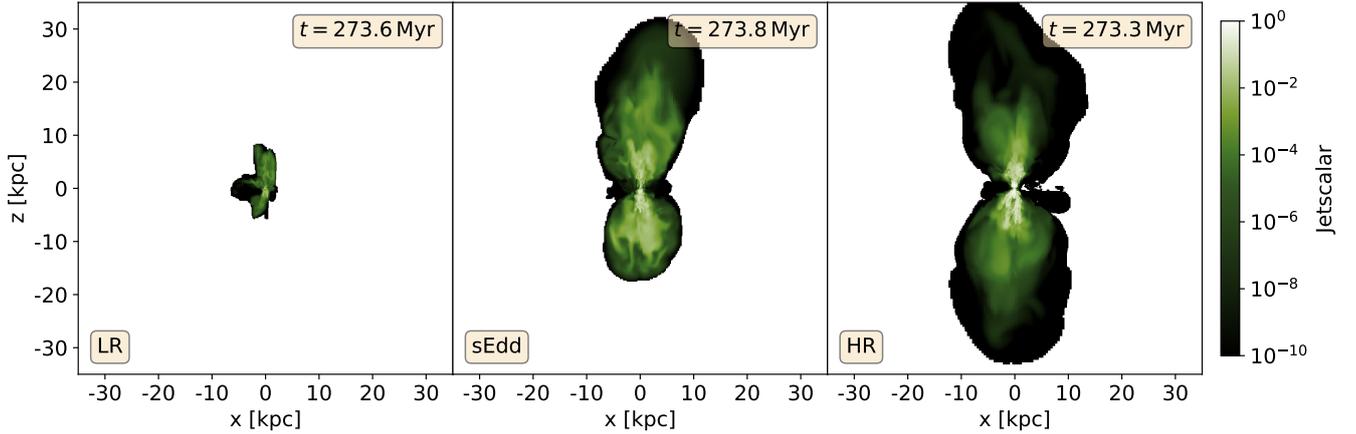

**Fig. 19.** *Left to right*: edge-on view at 274 Myr of the jet passive scalar for the 'LR', 'sEdd' and 'HR' runs. The jets extent tends to increase with resolution and has reached numerical convergence for spatial resolutions of order $\Delta x_{max} = 12$ pc (i.e. 'sEdd').

The BH mass evolution is shown in the left panel of Fig. 18. We find that generally with weaker super-Eddington jets, the BH grows more efficiently. However, long-term super-Eddington mass growth is not supported even with weak jets. 'sEdd_0.25' (blue circles) and 'sEdd_0.1' (green triangles) both have BHs growing below the Eddington limit (red dashed), with their jets powerful enough to regulate their growth (Regan et al. 2019). On the other hand, 'sEdd_0.05' (red squares) has a BH growing above the limit for a long period of time, due to the much weaker jets produced. This growth is a consequence of the gas in the accretion region which is not depleted by the outflows as effectively when little momentum is injected by the kinetic feedback. The 'sEdd_0.05' BH grows somewhat more than the BH in the 'sEddThm' simulation because of slightly denser gas in the equatorial direction and less dense gas in the polar one, likely caused by the anisotropic jets. The super-Eddington episodes still create strong changes in the gas temperature, but the effect is lessened when the jet efficiency $\eta_{jet}$ (or the MADness $f_{MAD}$ of the disc) is lower. For instance, 'sEdd' has temperature peaks at $3 \times 10^9$ K whilst the jets in 'sEdd_0.05' only increase the temperature up to $10^8$ K. This allows for more gas to be accreted and positively impacts BH growth.

In addition to the growth, we find that the fraction of time spent in the super-Eddington regime increases with decreasing disc MADness. As seen in the right panel of Fig. 18, the BH can spend up to 18% of its time in the super-Eddington regime for the 'sEdd_0.05' run, and converges down to a few percent with the 'sEdd_0.25' simulation. This further confirms that stronger jets decrease the frequency of super-Eddington episodes.

Galactic scale outflows are only found for 'sEdd_0.25', 'sEdd_0.1' and 'sEdd', which all have BHs growing on average below the Eddington limit. These large scale outflows are less frequent with the weaker jet feedback, as jets for lower MADness are lighter (in our algorithm) and therefore less able to punch through the dense galactic gas. We also find that weaker jets do not eject gas as far away as the more powerful ones, and tend not to stay collimated for long periods of time.

In conclusion, we show that the super-Eddington kinetic feedback, if weak, can allow for BHs to grow at a rate close to or slightly above the Eddington limit over long periods of time. Weak super-Eddington jets can be achieved by a combination of low spin $a$ and low MADness factor $f_{MAD}$, and in this study we focus on changing the disc MADness only. We do not see any evidence for strongly super-Eddington accretion, only mildly super-Eddington mass evolution for very low $f_{MAD}$. Finally, weaker jets produce less frequently galactic scale outflows. They often mix with their surroundings and do not stay collimated for very long.

### 4.4. Effect of the resolution

In addition to the previous analysis, we perform a series of resolution studies on the same halo, with high ('HR') and low ('LR') resolutions, respectively $\Delta x_{max}$ = 6 and 25 pc (see Table 2 for more details). To make sure that the study is solely focused on resolution, we use the initial conditions detailed in Table 1, at 140 Myr. We then switch the resolution, either increasing or decreasing it by a refinement level, let the galaxy reach an equilibrium state for ~20 Myr and then add a BH of $10^6$ $M_\odot$. This change can be visually seen in Fig. 9, with the galactic disc being thinner the higher the resolution, as well as SN-driven winds expanding farther out. After an Eddington-limited growth, we allow for super-Eddington accretion at around 200 Myr. For the 'LR' run, super-Eddington is allowed at $t$ = 199.8 Myr, a few Myr earlier than the 'HR' ($t$ = 206.1 Myr) and the 'sEdd' ($t$ = 206.4 Myr) simulations. Those different starting times of the super-Eddington regime are chosen so that the accretion rates around the BH are similar (between $20-50$ $\dot{M}_{Edd}$), making the comparison possible.

We follow an analysis similar to Sects. 4.1 and 4.2 and find very little differences between the three simulations. When increasing the resolution, we find that super-Eddington jets extended further and stay collimated for a longer period of time.

Within ~280 Myr, the jets reached 8 kpc in the 'LR' simulation, while converging to ~30 kpc for the 'sEdd' and 'HR', as shown in Fig. 19. In fact, the smaller the area $A$ of the jet at injection, the larger the final extent of the jet, as its velocity $v_{jet} \propto A^{-1}$. Within the studied timeframe, the BH is self-regulated due to the super-Eddington kinetic outflows. As can be seen in Fig. 20, the three BHs reach a similar mass, ~$3 \times 10^6 \, M_\odot$. We note that after 230 Myr, the 'HR' BH sees a sharper mass increase than the other two BHs. This is caused by the central region of the galaxy, which is denser ($\bar{\rho} \simeq 10^3$ cm$^{-3}$) by 1 dex than the other runs, and allows for larger mass growth. Increasing the resolution favors the creation of dense clumps inside the galaxy, which swirl in towards the BH and drive faster mass growth. However, they are insufficient to sustain average mass growth even close to the Eddington limit, as the BH reaches a mass of $3.2 \times 10^6 \, M_\odot$ in 70 Myr, only 10% more massive than the fiducial 'sEdd' BH.

Besides these small differences, the BH produces feedback of similar luminosity and the fraction of time spent in the super-Eddington regime is less than 1% in all three runs. We conclude that the behavior the BH growth with respect to the super-Eddington regime is robust against a change of resolution within the range of scales tested here, of the order of those reached in the latest high resolution cosmological simulations (e.g. Dubois et al. 2021). Since this isolated setup is not conducive to study spatial resolutions 0.1–1 kpc, we cannot assess whether the effect of super-Eddington AGN feedback is correctly captured at low resolution or, e.g., a boost factor would be needed (Booth & Schaye 2009; DeGraf et al. 2017; Chabanier et al. 2020). We postpone such investigation to future work using cosmological simulations.

## 5. Discussion

One important underlying caveat of the work presented here are the assumptions made in how the jet is modelled. Ideally, one would like to inject the jet energy at the scale of the BH accretion disc, i.e. on length-scales of the order of the Schwarzschild radius of the BH, $r_{Sch}$. Given that $r_{Sch} \approx 10^{-7}$ pc for a ~$10^6 \, M_\odot$, this is technically unfeasible for the galaxy-scale study presented here. As a result, one needs to make a choice on how the jet propagates from the Schwarzschild scale to the resolution scale of the simulation, here $\Delta x = 12$ pc.

In this paper, we make the assumption that the kinetic energy is conserved, but that the jet slows down from $c$ at the Schwarzschild scales to $0.1c$ at resolution scales (see Sect. 2 for details) due to mass entrained as it propagates. We assume that the mass entrained by the jet is not directly removed from the accretion flow, which is unaffected, but instead swept up on scales between the accretion flow and the resolution of our simulation. This is reasonable due to the 8 orders of magnitude of separation of scales between the physical size of the BH and the simulation resolution. GRMHD simulations exceeding the Eddington limit (e.g. Jiang et al. 2014; McKinney et al. 2014; Sądowski et al. 2014) find that outflows on scales from ~10 to ~100 Schwarzschild radii (far below our resolution scale), can carry away most of the inflowing mass, but on scales up to $10^4$ Schwarzschild radii, the mass loading of the jet can increase by an order of magnitude for a smooth medium even in the sub-Eddington case (Chatterjee et al. 2019), and perhaps more for a clumpy medium or larger scales. As our numerical injection scale is 8 orders of magnitude above the physical injection scale of BH jets, our mass loading factor of $\beta = 0.32$–32 (see Table 2) seems consistent with previous work. This increase in jet mass is only necessary due to the gap in scales between the simulation and the BH. If we were able to inject the jet at the speed of light, $c$, at scales comparable to the Schwarzschild radius, no such increase in jet mass loading would be required as we would be able to self-consistently split the accreted mass into BH mass growth and outflow.

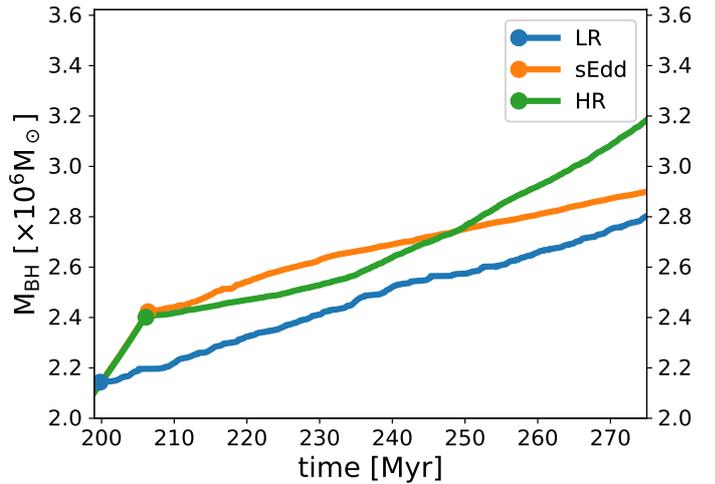

**Fig. 20.** Evolution of the BH mass for the 'LR' (solid blue), 'sEdd' (solid orange) and 'HR' (solid green) simulations. Self-regulation is reached in the three cases in very similar fashion.

Other choices would have been possible, such as assuming that the extra mass in the jet comes from the accretion flow, which would mean reducing the mass growth rate onto the BH, and resulting jet power, by a factor of $1/(\beta_{jet} + 1)$. This choice, applied to the super-Eddington regime, would more accurately reflect mass loss in the accretion disc via winds found in GRMHD simulations (Sądowski et al. 2013, 2014). However, we remind the reader that since this model is implemented for cosmological simulations, accretion has to be boosted to meet observational constraints, meaning that an additional ad hoc factor (Booth & Schaye 2009) would probably be needed to compensate for this suppression.

## 6. Conclusions

In this paper, we examine the impact of including the super-Eddington regime on BH growth and on the gas properties from 10 pc to several kpc scales, in an isolated galaxy at $z = 4$. A BH with initial mass $M_{BH} = 10^6 \, M_\odot$ is placed in the centre of a $M_* \simeq 10^9 \, M_\odot$ galaxy where cold and dense gas can trigger super-Eddington episodes. To study this regime, a modification of the RAMSES algorithm was made, to remove the cap at the Eddington limit and implement both types of feedback (kinetic and thermal) that are believed to characterise super-Eddington sources (Sect. 2). Our main findings are as follows.

- The impact of super-Eddington AGN feedback on SF is minimal due to the narrow outflows launched by the BH (Sect. 4.1.1).
- With super-Eddington AGN feedback the BH reaches self-regulation within a few Myr (depending on the type of AGN feedback), even in an environment favorable for super-Eddington accretion. Sharp and instantaneous drops down to sub-Eddington regimes occur after each super-Eddington event that peak at around 2–3 times the Eddington limit (Sect. 4.1.2).
- The formation of super-Eddington jets reduces the amount of time spent in the super-Eddington regime (<1% in 'sEdd',

compared to ∼10% in 'sEddThm' without jets) and shortens the time before self-regulation is reached. The more often a BH is in the super-Eddington regime, the more mass is accumulated in this regime (up to ∼70% without jets in 'sEddThm'). Including jets in the super-Eddington regime is therefore necessary to capture its properties and effects (Sect. 4.2.1).

– On small scales (10–100 pc), super-Eddington AGN feedback heats up the gas to $10^{8-9}$ K instantaneously. The powerful jets eject most of the gas outside of the galaxy, reducing the gas density around the BH significantly; whilst the weak thermal feedback creates a hot "bubble" without sufficient momentum to launch gas further than the BH accretion region. Gas gets replenished by infalls within a few kyr, allowing for another super-Eddington episode to be triggered (Sect. 4.2.2).

– On larger scales (∼kpc), super-Eddington jets do not significantly impact gas inflows. As long as the interstellar medium is not too dense, if the galaxy disc and the jets are orthogonal, the outflows escape the galaxy with a steady ∼$10^3$ km s$^{-1}$, stay collimated and create a path for subsequent super-Eddington jets (Sect. 4.2.3).

– Lower super-Eddington jet efficiencies, occurring if the BH spin is low or if the MADness state (magnetic saturation) of the accretion disc is weak, allow for more frequent super-Eddington events and for more significant BH growth, since accretion is not suppressed as often or as strongly as for powerful jets (Sect. 4.3).

– Even with the lowest jet feedback efficiencies, there is no evidence of strong super-Eddington growth: BHs can only grow slightly above a BH that would continuously accrete at Eddington (∼15% more in BH mass; Sects. 4.2.1 and 4.3).

This paper is a first step at studying the impact on BH growth by the super-Eddington regime of AGN feedback, within a realistic galaxy environment. Cosmological simulations are needed to understand whether the super-Eddington regime can explain the billion solar masses of BHs powering the most luminous quasars in the high-redshift Universe. Our findings suggest that the right combination of BH spin and super-Eddington AGN feedback strength may lead to a window of opportunity for mildly super-Eddington mass evolution.

*Acknowledgements.* The authors thank Hugo Pfister, Maxime Trebitsch and Pierre Cox for insightful discussions. R.S.B. gratefully acknowledges funding from Newnham College, University of Cambridge and the ANR grant LYRICS (ANR-16-CE31-0011). This work was granted access to the HPC resources of CINES under the allocations A0080406955 and A0100406955 made by GENCI. This work has made use of the Infinity Cluster hosted by Institut d'Astrophysique de Paris. We thank Stéphane Rouberol for smoothly running this cluster for us. Visualisations in this paper were produced using the YT PROJECT (Turk et al. 2011).

## Appendix A: Comparison with $v_{jet} = c$

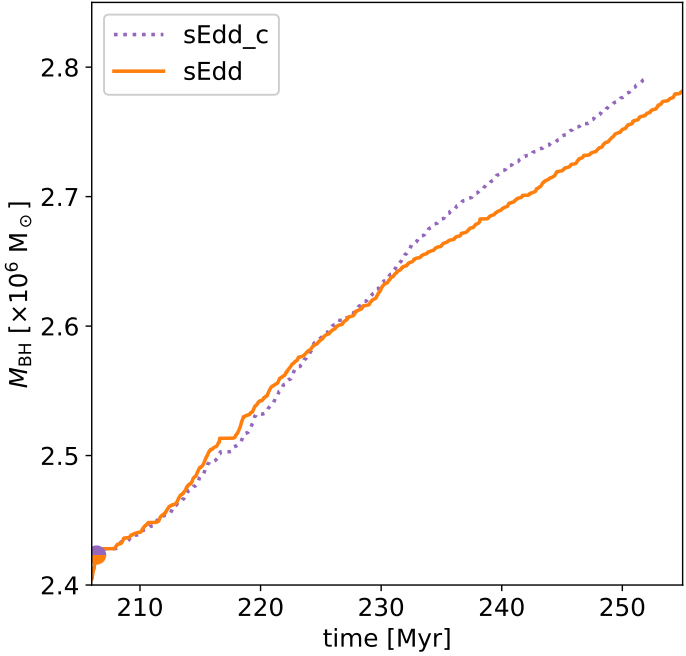

**Fig. A.1.** BH mass evolution until $t \simeq 250$ Myr for the 'sEdd_c' (dashed violet) and 'sEdd' (solid orange) simulations.

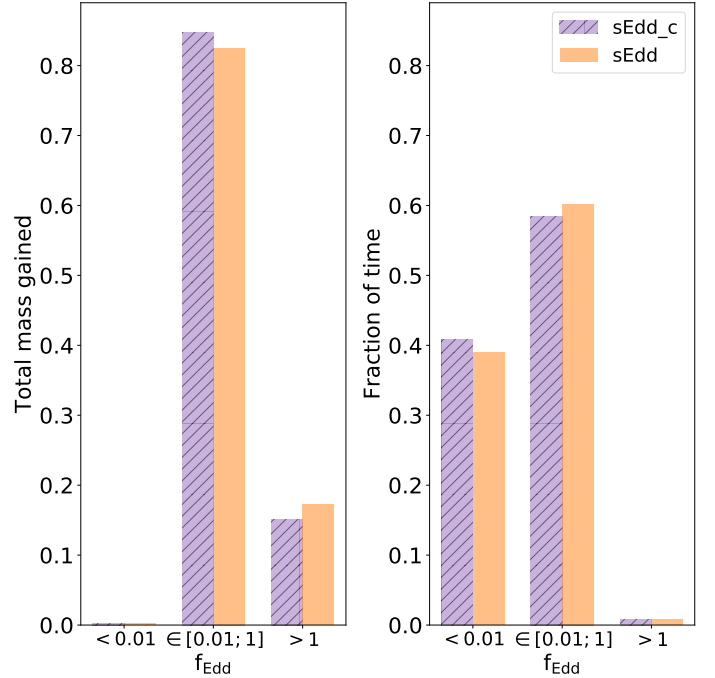

**Fig. A.2.** Fractional values of the total mass gained (left) and fraction of time spent (right) by the BH in the 'sEdd_c' (hashed violet) and 'sEdd' (filled orange) simulations in $f_{Edd}$ bins, mirroring the three accretion/feedback regimes, from the moment super-Eddington started ($t \simeq 206.4$ Myr) until $t \simeq 250$ Myr.

To test the impact of our choices on the way BH accretes and how the jet is loaded in our simulations, we performed an additional simulation called 'sEdd_c'. It is identical in all points to our fiducial simulation 'sEdd' besides the jet velocity, which we set $v_{jet} = c$, giving a mass loading factor $\beta_{jet} = 0.32 \sim 1$. In this new simulation, we therefore assume that the numerical injection scale is the same as the physical injection scale.

Thanks to this new simulation, we are able to have $\dot{M}_{jet} = 0.32 \dot{M}_{BH} = 0.32 \dot{M}_{BHL}$, meaning that the outflow rates are of the order of $\dot{M}_{BHL}$. We stress that the mass accreted onto the BH follows the same formalism as before, i.e. $\dot{M}_{BH} = (1 - \epsilon_r)\dot{M}_{acc} = (1 - \epsilon_r)\min(\dot{M}_{BHL}, \dot{M}_{floor})$. In Fig. A.1, we compare the final BH mass, and in Fig. A.2, the mass accreted in each regime, between 'sEdd_c' and the fiducial 'sEdd' simulations, in order to see if changing the outflow velocity, and subsequently mass loaded in the jet, makes a difference to the BH growth. In both simulations, for a given $\dot{M}_{BH}$, the energy injected in the BH surroundings is the same ($\propto \eta_{jet}\dot{M}_{BH}c^2$); however the momentum differs, since it is $\propto \beta_{jet}\dot{M}_{BH}v_{jet}$. For instance, it corresponds to $3.2\dot{M}_{BH}c$, for 'sEdd'; and $0.32\dot{M}_{BH}c$, for 'sEdd_c'.

By the end of the simulation, the final BH mass is very similar in both cases; the total mass accreted in each regime, as well as the time spent in each regime, are also approximately the same. We conclude that loading the jet with either $v_{jet} = 0.1c$ giving $\beta_{jet} = 32$ (fiducial, 'sEdd') or with $v_{jet} = c$ giving $\beta_{jet} = 0.32$ (faster but less loaded jet, 'sEdd_c') does not lead to significant differences regarding our conclusions on BH self-regulation.